\documentclass{ws-procs9x6}

\begin{document}

\title{Light mesons and infrared behavior of the running coupling
constant in QCD}

\author{M. Baldicchi and G. M. Prosperi}

\address{Dipartimento di Fisica dell'Universit\`a di Milano; I.N.F.N.
Sez. di Milano;\\ via Celoria 16,
I20133 Milano, Italy; E-mail: prosperi@mi.infn.it}


\maketitle

\abstracts{A previous method for handling bound states in QCD is briefly
revisited. Taking advantage of the Feynman-Schwinger representation for the
iterated quark propagator in an external field, it is possible to give
closed representations for certain appropriate (second order) two point
and four point Green functions, $H^{(2)}(x-y)$ and $H^{(4)}(x_1,x_2,y_1,y_2)$,
as path integrals on quark world lines. Then, starting from
reasonable assumptions on the Wilson line correlators, a Bethe-Salpeter
equation for $H^{(4)}$ and a Dyson-Schwinger equation for $H^{(2)}$ can
be obtained, which are consistent with the Goldstone theorem in the chiral
limit. Such equations are too complicate to be solved directly. However, a
reduced Salpeter equation can be derived which is tractable and has been
applied to a calculation of the meson spectrum. The results are in general
good agreement with the data, but with the important exceptions of the light
pseudo scalars (that are related to the breaking of the chiral symmetry).
    In this scenario two important improvements can be introduced: a) the
fixed coupling constant can be replaced by a running coupling constant
$\alpha_{\rm s}(Q^2)$ appropriately
modified in the infrared region; b) the fixed mass in
the reduced equation can be replaced for light quarks by an effective
mass depending on the momentum of the particle, as suggested by the form of
the DS equation. Then even the light pseudo scalar mesons can be made
to agree with to their experimental value.}


\section{Introduction}

  In previous papers we have introduced what we have called a second order
Bethe-Salpeter formalism, which works in terms of certain appropriate four
point and two point Green functions, $ H^{(4)}(x_1, x_2, y_1, y_2) $ and
$ H^{(2)} (x-y) \, $\cite{bmp}. Taking advantage of a Feynman-Schwinger
representation for the ``second order'' quark propagator in an external
field, it is possible to write $ H^{(2)}$ and
$ H^{(4)}$ as path integrals on quark or antiquark world
lines joining $y$ to $x$, $y_{1,2}$ to $x_{1,2}$.
In such representations the gauge field appears only trough {\it Wilson
line } correlators like ${1\over\sqrt 3}{\rm Tr_{color}} \langle
\exp [i \int_{y}^{x} dz^\mu A_\mu (z)]$ or ${1\over 3}{\rm Tr_{color}}
\langle \exp [i \int_{y_1}^{x_1} dz^\mu A_\mu (z)]
\exp [i \int_{x_2}^{y_2} dz^\nu A_\nu (z)]  \rangle $
which reduce to ordinary Wilson loops $W$ in the limit
$ x \to y $ or $ x_1 \to x_2$ and $y_1 \to y_2$. In analogy with a
usual assumption on $W$ such correlators are written as the sum of their
perturbative expressions and appropriate area terms and it is possible
to obtain a confining Bethe-Salpeter equation for the quantity $H^{(4)}$
and a corresponding Dyson-Schwinger equation for $H^{(2)}$.
 
   The above equations are too complicate to be solved directly. However,
by a conventional three dimensional reduction, one can derive a tractable
squared mass operator that can be applied to an evaluation of the
spectrum~\cite{simon}.
With an appropriate choice of the parameters (quark
masses, coupling constant, string tension) the results can be made
in good agreement with the data in the entire framework of the
light-light, light-heavy and heavy-heavy quark-antiquark sectors,
with the important exception, however, of the light pseudo scalar
mesons \cite{quadratic}.

  In this paper we want to discuss two important improvements to the
above scenario:

  a) the fixed coupling constant $\alpha_{\rm s}$ is replaced by a running
coupling constant $\alpha_{\rm s}(Q^2)$, which amounts to take into account
higher order contributions in $\alpha_{\rm s}$ (and in particular closed
quark loops);

  b) an effective mass for the light quarks is introduced which depends on
the momentum of the particle, as suggested by the form of the DS equation.

  As well known, the usual perturbative expressions for $\alpha_{\rm s}(Q^2)$
have an unphysical singularity for $ Q^{2} = \Lambda^{2}_{\rm QCD} $, which
would be disastrous for our purpose. However, various modification
have been proposed for the infrared region \cite{sanda,shirkov}.
We have considered in particular
the Shirkov-Solovtsov prescription, which rests only on general
analyticity requirements. We find that with such
prescription, and parametrazing the effective squared light quark masses
by a simple polynomial, even the light pseudo scalar mesons can be made
in agreement with their experimental values. It should be stressed that
in the fit only the quark masses are
treated as free parameters, while the constant $\Lambda$ and $\sigma$
occurring in the BS kernel are a priori fixed on the basis of high energy
phenomenology and lattice simulations. Results in this line
were already published in ref. [\cite{infrared}]; here we present
a more systematic study.
   
  The remaining part of the paper is organized as the following plan: in
sections 2 and 3 we briefly revue the second order Bethe-Salpeter formalism,
in section 4 we discuss the running coupling constant and effective mass,
in sections 5 and 6 we reports the results and make some conclusions.


\section{Second order correlators}

After integrating out the fermionic fields,
the appropriate ordinary (first order) four point function for the
quark-antiquark bound state problem
can be written
\begin{eqnarray}
G^{(4)}(x_1,x_2;y_1,y_2) &=&
\frac{1}{3}\sum_{ab}\langle 0 | {\rm T} 
\psi_{1a}(x_1)
\overline{\psi}_{2a}(x_2)
\overline{\psi}_{1b}(y_1) \psi_{2b}(y_2)
|0\rangle =
\nonumber \\
&=& - \frac{1}{3} {\rm Tr_{Color}} \langle S_1(x_1,y_1;A) 
S_2(y_2,x_2; A) \rangle ,
\label{eq:fo4point}
\end{eqnarray}
%
%
%
\noindent
Here $ a $ and $ b $ are color indexes,
the subscripts 1 and 2 refer to the quark and the antiquark
respectively, projection on the color singlet has been performed,
$ S(x,y;A) $  denotes the quark propagator in an external field,
\begin{equation}
( i\gamma^\mu D_\mu -m) S(x,y;A) =\delta^4(x-y)\,,
\label{eq:foprop}
\end{equation}
\noindent
and
\begin{equation}
\langle f[A] \rangle = \int  DA\, M_F [A]\, e^{iS_G[A]} f[A]  \,,
\label{eq:expt}
\end{equation}
\noindent
with
$$
M_F[A] = {\rm Det} \Pi_{j=1}^2 [1 + g\gamma^\mu A_\mu
( i\gamma_j^\nu \partial_{j\nu} - m_j)^{-1}].
$$


    The {\it second order} four point function is defined by
\begin{equation}
H^{(4)}(x_1,x_2;y_1,y_2)= -{1\over 3} {\rm Tr _{color}}
\langle \Delta_1^\sigma (x_1,y_1;A)
{\Delta}_2^\sigma (y_2,x_2;A)\rangle \, ,
\label{eq:so4point}
\end{equation}
\noindent
where $\Delta_1^\sigma (x,y;A)$ is the  {\it second order} propagator that
satisfies the second order differential equation
\begin{equation}
(D_\mu D^\mu +m^2 -{1\over 2} g \, \sigma^{\mu \nu} F_{\mu \nu})
\Delta^\sigma (x,y;A) = -\delta^4(x-y) \, ,
\label{eq:soprop}
\end{equation}
($\sigma^{\mu \nu} = {i\over 2} [\gamma^\mu, \gamma^\nu]$) and is related to
$S(x,y;A)$ by
\begin{equation}
S(x,y;A) = (i \gamma^\nu D_\nu + m) \Delta^\sigma(x,y;A) \,.
\label{eq:fotoso}
\end{equation}
  The quantity $H^{(4)}$ is related to $G^{(4)}$ by an integro-differential
operator, of the type
$[(i\gamma_1^\mu \partial_{1\mu}+m_1)
(i\gamma_2^\nu \partial_{2\nu}+m_2)+\dots] \,, $
which we do not need to specify in detail. The important fact is that the two
functions are completely equivalent for the determination of the bound states
since their Fourier transforms $\hat H^{(4)}$ and $\hat G^{(4)}$
have related analyticity properties and the same poles.

   The advantage in considering second order quantities is that it is
possible to write for $\Delta^\sigma(x,y;A)$ a generalized Feynman-Schwinger
representation, i. e. to solve eq. (\ref{eq:soprop}) in terms of a quark
path integral
\begin{eqnarray}
\Delta^\sigma (x,y; A) &=&  -{i \over 2} \int_0^\infty ds \int_y^x
   Dz \exp [-i \int_0^s d\tau {1\over 2} (m^2
   +\dot z ^2)] \nonumber \\
  & & \qquad \qquad S_0^s {\rm P} \exp[ ig \int_0^s d \tau
   \dot z ^\mu A_{\mu}(z)] \, ,
\label{eq:fsrepr}
\end{eqnarray}
where the world-line $z^\mu=z^\mu(\tau)$ connecting $y$ to $x$ is written
in the four-dimensional
language in terms of an additional parameter $\tau$,
$ S_0^s = {\rm T} \exp \Big [ -{1\over 4} \int_0^s d \tau
 \sigma^{\mu \nu} {\delta \over \delta S^{\mu \nu}(z)}
\Big ] $ and $ \delta
S^{\mu \nu} = dz^\mu \delta z^\nu - dz^\nu \delta z^\mu $ (the functional
derivative being defined through an arbitrary deformation, $ z \rightarrow
z + \delta z $, of the world-line).

  Replacing  eq. (\ref{eq:fsrepr}) in  eq. (\ref{eq:so4point}) a similar
representation can be obtained for the 4-point function
\begin{eqnarray}
& &H^{(4)}(x_1,x_2;y_1,y_2)  = ({1 \over 2})^2 \int_0^{\infty} d s_1
\int_0^{\infty} d s_2
 \int_{y_1}^{x_1}  Dz_1\int_{y_2}^{x_2} D z_2
\nonumber \\
& & \qquad \qquad \exp \Big \{ -{i \over 2}
 \int_0^{s_1} d\tau_1 (m_1^2 +\dot{z}_1^2) - {i\over 2} \int_0^{s_2}
d\tau_2 (m_2^2 +\dot{z}_2^2)\Big \}
 \label{eq:fs4point} \\
& & {1\over 3} S_0^{s_1}  S_0^{s_2} {\rm Tr _{color}}
\langle  {\rm P}  \exp \big \{ig\int_{y_1}^{x_1} dz_1^{\mu} \, A_{\mu}(z_1)
\big \}
 {\rm P}  \exp \big \{ig\int_{x_2}^{y_2} dz_2^{\mu} \, A_{\mu}(z_2)
\big \}
\rangle \, ,
\nonumber
\end{eqnarray}
The interesting aspect of the above equation is that the gauge field
appears in it only through the expectation value of the product of the two
Wilson lines.

    Similarly the second order two point function (uncolored full
quark propagator) $H^{(2)}(x-y) = {i \over \sqrt{3}}{\rm Tr_{color}}
\langle S(x,y:A)\rangle$  can be written
\begin{eqnarray}
H^{(2)}(x-y)  &=& {1 \over 2} \int_0^{\infty} d s
 \int_y^x  Dz
 \exp \big \{  -{i \over 2}
 \int_0^s d\tau (m^2 +\dot{z}^2) \big \} \nonumber \\
& &  {1 \over \sqrt{3}} S_0^s {\rm Tr _{\rm color}}
\langle  {\rm P}  \exp \big \{ig\int_y^x dz^\mu \, A_\mu(z)
\big \}
\rangle \, .
\label{eq:fs2point}
\end{eqnarray}

     Notice that, for large separations,
we can also write in operatorial form
%
\[
H^{(4)}(x_1,x_2;y_1,y_2) =
\frac{1}{3}\sum_{ab}\langle0|{\rm T} 
\phi_{1a}(x_1)
\overline{\psi}_{2a}(x_2)
\overline{\psi}_{1b}(y_1) \phi_{2b}(y_2)
|0\rangle
\label{eq:op4point}
\]

and

\[
H^{(2)}(x-y) = \sum_a\langle0|{\rm T}\phi_a(x)
\overline{\psi}_a (y)|0\rangle \, ,
\]

\noindent
$\phi (x)$ being defined by

\[
\psi(x) = (i \gamma^\nu D_\nu + m) \phi(x) \, .
\]
%
The above forms are important for a definition of the relativistic wave
function.


\section{Bethe-Salpeter and Dyson-Schwinger equations}

  In the limit $x_2 \to x_1$, $y_2 \to y_1$ or $y \to x$ the two Wilson lines
occurring in (\ref{eq:fs4point}) or the single line occurring in
(\ref{eq:fs2point}) close in a single Wilson loop $\Gamma$
\begin{equation}
W = \langle {1 \over 3} {\rm Tr_{\rm color} P}\exp \big[ \oint_\Gamma dz^\mu
A_\mu(z) \big] \rangle\,.
\label{eq:wdef}
\end{equation}
  As in the previous papers we assume that in a first approximation
$i\ln W$ can be written as the sum of its perturbative expression and an
area term
$
i\ln W = (i\ln W)_{\rm pert} + \sigma S\,.
$
Then, at the lowest order in the coupling constant, we can assume
\begin{eqnarray}
&& i\ln W = {4\over 3} g^2 \oint dz^\mu \oint dz^{\nu \prime}
D_{\mu \nu}(z-z^\prime) +
\label{eq:wilson} \\
&& \sigma \oint dz^0 \oint dz^{0 \prime} \delta (z^0-z^{0\prime})
|{\bf z} - {\bf z}^\prime| \int_0^1 d\lambda
 \Big \{ 1 -  [\lambda {d{\bf z}_{\rm T} \over dz^0}
 + (1-\lambda) {d{\bf z}_{\rm T} ^\prime \over dz^{0 \prime}} ]^2 
\Big \}^{1\over 2} \, . \nonumber
\end{eqnarray}

    Notice that the surface term in (\ref{eq:wilson}) is written as
the algebraic sum of successive equal time straight strips
($ {\bf z}_{\rm T} $ denotes the transversal component of
$ {\bf z} $). For a flat loop or for other special geometries (e. g. for two
quarks uniformly rotating around their fixed center of mass)
this coincides obviously
with the plane or the minimum surface
delimited by $\Gamma$. That is not generally the case, indeed
the right hand side of (\ref{eq:wilson}) usually depends on the reference
frame. Since, however, in contrast e. g. with $S_{\rm min}$,
such quantity maintains many of the
analytic properties of the original $i \ln W$,
we shall assume (\ref{eq:wilson}) to be valid for an arbitrary loop
in the center of mass reference frame.
Actually we shall assume this even for
$x_2 \ne x_1$, $y_2 \ne y_1$ or $y \ne x$,
in analogy with what happens in the pure perturbative case.
In this way single perturbative and confinement contributions
are put on the same foot and we may refer to them as a {\it gluon
exchange} and a {\it string connection} between the two quarks.

  Replacing (\ref{eq:wilson}) in (\ref{eq:fs4point}) and (\ref{eq:fs2point})
we 
obtain the following equations
%
%
%
%
%
%
\begin{eqnarray}
& & H^{(4)}(x_1,x_2;y_1,y_2) = ({1 \over 2})^2 \int_0^{\infty}
   d s_1  \int_0^{\infty} d s_2
  \int_{y_1}^{x_1} Dz_1\int_{y_2}^{x_2}  D z_2 \nonumber \\
& &  \exp \Big\{  -{i \over 2}
  \sum_{j=1}^2  \int_0^{s_j} d\tau_j (m_j^2 +\dot{z}_j^2) \Big\}
  S_0^{s_1} S_0^{s_2}
  \exp \Big\{ i \sum_{j=1}^2 \int_0^{s_j} d \tau_j 
\label{eq:fsqq}
\\
& &
  \int_0^{\tau_j} d \tau _j ^\prime
    E( z_j -z_j^\prime ; \dot z _j , \dot z _j^\prime)
    -i \int_0^{s_1}
   d \tau_1  \int_0^{s_2} d \tau _2
   E( z_1 -z_2 ; \dot z _1 , \dot z _2) \Big\}
\nonumber
\end{eqnarray}
and
\begin{eqnarray}
H^{(2)}(x-y) &=& {1 \over 2} \int_0^{\infty} d s
 \int_y^x  Dz
 \exp \big \{  -{i \over 2}
 \int_0^s d\tau (m^2 +\dot{z}^2) \big \} \nonumber \\
& &  \qquad \qquad \qquad  S_0^s
\exp \big \{i \int_0^s \int_0^\tau
E(z - z^\prime; \dot z , \dot z ^\prime) \big \} \ ,
\label{eq:fsq}
\end{eqnarray}
where we have set
\begin{equation}
E(\zeta;p,p^\prime) = E_{\rm pert} (\zeta;p,p^\prime) +
      E_{\rm conf} (\zeta;p,p^\prime)
\label{eq:eea}
\end{equation}
with
\begin{equation}
\left \{
\begin{array}{ll}
E_{\rm pert}  = 4 \pi {4 \over 3} \alpha_{\rm s} D_{\mu \nu}(\zeta)
 p^\mu p^{\prime \nu}  \nonumber \\
E_{\rm conf}  = \delta (\zeta _0) \vert {\bf \zeta} \vert \epsilon (p_0)
 \epsilon (p_0^{\prime})
 \int_0^1 d \lambda \{ p_0^2 p_0^{\prime 2} - [\lambda p_0^{\prime}
 {\bf p}_{\rm T}
 + (1- \lambda) p_0 {\bf p}_{\rm T} ^{\prime} ]^2 \}^{1 \over 2} \, .
\end{array}
\right.
\label{eq:eeb}
\end{equation}
%
%
%
   From eqs. (\ref{eq:fsqq}) and (\ref{eq:fsq}),
by various manipulation
and using an appropriate iterative procedure, a Bethe-Salpeter equation
for the function $H^{(4)}(x_1,x_2;y_1,y_2)$ and a Dyson-Schwinger
equation for $H^{(2)}(x-y)$ can
be derived in a kind of generalized ladder and rainbow
approximation respectively
in the form
%
\begin{eqnarray}
H^{(4)}(x_1,x_2;y_1,y_2) &=& H_{1}^{(2)}(x_1-y_1) \, H_{2}^{(2)}(x_2-y_2) -
\qquad \qquad \qquad \qquad \quad
\nonumber \\
    & - & i \int d^4 \xi_1 d^4 \xi_2 d^4 \eta_1 d^4 \eta_2
    H_{1}^{(2)}(x_1-\xi_1) \, H_{2}^{(2)}(x_2-\xi_2)
\nonumber \\
    & & \quad \times I_{ab}(\xi_1,\xi_2;\eta_1,\eta_2)\,
    \sigma_1^a \,
    \sigma_2^b \, H^{(4)}(\eta_1,\eta_2;y_1,y_2) \, ,
\label{eq:bseq}
\end{eqnarray}
and
\begin{eqnarray}
  H^{(2)}(x-y) &=& H_{0}^{(2)}(x-y) + i \int d^4 \xi
  d^4 \eta d^4 \xi^\prime
       d^4 \eta^\prime H_{0}^{(2)}(x-\xi)
\qquad \qquad \qquad
\nonumber \\
  & & 
\qquad
\times I_{ab}(\xi,\xi^\prime;\eta,\eta^\prime)
   \sigma^a H^{(2)}(\eta - \eta^\prime) \sigma^b
   H^{(2)}(\xi^\prime - y) \, ,
\label{eq:sdconf}
\end{eqnarray}
where 
$a, \, b = 0, \, \mu\nu$, we have set $\sigma^0=1$,
and $ H_{1}^{(2)} $ and
$ H_{2}^{(2)} $ denote the second order full
quark and the antiquark propagators respectively.

   In momentum representation, the corresponding homogeneous
BS-equation can be written
\begin{eqnarray}
\Phi_P (k) &=& -i \int {d^4u \over (2 \pi)^4}
   \hat I_{ab} \big (k-u, {1 \over 2}P
   +{k+u \over 2},
   {1 \over 2}P-{k+u \over 2} \big )\nonumber \\
    & & \qquad \qquad
   \hat H_{1}^{(2)}   \big ({1 \over 2} P  + k \big )
      \sigma^a  \Phi_P (u) \sigma^b
   \hat H_{2}^{(2)} \big (-{1 \over 2} P + k \big ) \, ,
\label{eq:bshom}
\end{eqnarray}
\noindent
the center of mass frame has to be understood, $P=(m_B, {\bf 0})$
and $\Phi_P (k)$ denotes the appropriate {\it second order} wave function
\[
  \langle 0|\phi({\xi \over 2}) \bar\psi(-{\xi \over 2}) |P\rangle  =
  {1\over (2\pi)^2} \Phi_P (k) e^{-ik\xi} \,.
\]
%
  Similarly, in terms of the irreducible self-energy, defined by
$$\hat H^{(2)}(k) ={i\over k^2-m^2} + {i\over k^2-m^2} i\Gamma (k)
\hat H^{(2)}(k) \, , $$
the DS-equation can be written also
\begin{equation}
\hat \Gamma(k) =  \int {d^4 l \over (2 \pi)^4}  \,
\hat I_{ab} \Big ( k-l;{k+l \over 2},{k+l \over 2} \Big )
\sigma^a \hat H^{(2)}(l) \, \sigma^b \ .
\label{eq:dshom}
\end{equation}
   The kernels in
(\ref{eq:bshom}) and (\ref{eq:dshom})
are the same in the two equations, consistently with the
requirement of chiral symmetry limit \cite{chiral}, and are given by
\begin{eqnarray}
& & \hat I_{0;0} (Q; p, p^\prime)  =  4 \int d^4 \zeta \, e^{iQ \zeta}
   E(\zeta ;p,p^\prime) =
   16 \pi {4 \over 3} \alpha_{\rm s} p^\alpha p^{\prime \beta}
  \hat D_{\alpha \beta} (Q)  + \nonumber \\
& &  \quad + 4 \sigma  \int d^3 {\bf \zeta} e^{-i{\bf Q}
   \cdot {\bf \zeta}}
    \vert {\bf \zeta} \vert \epsilon (p_0) \epsilon ( p_0^\prime )
   \int_0^1 d \lambda \{ p_0^2 p_0^{\prime 2} -
   [\lambda p_0^\prime {\bf p}_{\rm T} +
   (1-\lambda) p_0 {\bf p}_{\rm T}^\prime ]^2 \} ^{1 \over 2} \nonumber \\
& & \hat I_{\mu \nu ; 0}(Q;p,p^\prime) = 4\pi i {4 \over 3} \alpha_{\rm s}
   (\delta_\mu^\alpha Q_\nu - \delta_\nu^\alpha Q_\mu) p_\beta^\prime
   \hat D_{\alpha \beta}(Q)  - \nonumber \\
& & \qquad \qquad \qquad  - \sigma  \int d^3 {\bf \zeta} \, e^{-i {\bf Q}
\cdot {\bf \zeta}} \epsilon (p_0)
   {\zeta_\mu p_\nu -\zeta_\nu p_\mu \over
   \vert {\bf \zeta} \vert \sqrt{p_0^2-{\bf p}_{\rm T}^2}}
   p_0^\prime  \nonumber \\
& & \hat I_{0; \rho \sigma}(Q;p,p^\prime) =
   -4 \pi i{4 \over 3} \alpha_{\rm s}
   p^\alpha (\delta_\rho^\beta Q_\sigma - \delta_\sigma^\beta Q_\rho)
   \hat D_{\alpha \beta}(Q) + \nonumber \\
& & \qquad \qquad  \qquad  + \sigma  \int d^3 {\bf \zeta} \, e^{-i{\bf Q}
  \cdot {\bf \zeta}} p_0
  {\zeta_\rho p_\sigma^\prime - \zeta_\sigma p_\rho^\prime \over
  \vert {\bf \zeta} \vert \sqrt{p_0^{\prime 2}
   -{\bf p}_{\rm T}^{\prime 2}} }
  \epsilon (p_0^\prime)  \nonumber \\
& & \hat I_{\mu \nu ; \rho \sigma}(Q;p,p^\prime) =
   \pi {4\over 3} \alpha_{\rm s}
  (\delta_\mu^\alpha Q_\nu - \delta_\nu^\alpha Q_\mu)
  (\delta_\rho^\alpha Q_\sigma - \delta_\sigma^\alpha Q_\rho)
  \hat D_{\alpha \beta}(Q) \, ,
\label{eq:imom}
\end{eqnarray}
\noindent
where in the second and in the third equation $\zeta_0 = 0$ has to be
understood.

  To  find the $q \overline q $ spectrum, in principle one should solve first
(\ref{eq:dshom}) and use the resulting propagator in (\ref{eq:bshom}). In
practice this turns out to be a difficult task and one has to resort to the
three dimensional equation which can be obtained from (\ref{eq:bshom}) by the
so called instantaneous approximation. This consists
in replacing $\hat H_{j}^{(2)}(k)$ in (\ref{eq:bshom}) with the free
quark propagator ${i \over k^2 -m_j^2}$ and the kernels
$\hat{I}_{ab}$ with its so called instantaneous approximation
$ \hat{I}_{ab}^{\rm inst}({\bf k}, {\bf k}^\prime)$ \cite{bmp}.

  The reduced equation takes the form of the eigenvalue equation  for a
squared mass operator
$ M^2 = M_0^2 + U $,
with  $ M_0 = \sqrt{m_1^2 + {\bf k}^2} + \sqrt{m_2^2 + {\bf k}^2} $ and
\begin{equation}
   \langle {\bf k} \vert U \vert {\bf k}^\prime \rangle =
        {1\over (2 \pi)^3 }
        \sqrt{ w_1 + w_2 \over 2  w_1  w_2}\, \hat I_{ab}^{\rm inst}
        ({\bf k} , {\bf k}^\prime)  \sqrt{ w_1^\prime + w_2^\prime \over 2
         w_1^\prime w_2^\prime}\sigma_1^a \sigma_2^b \,
\label{eq:quadrrel}
\end{equation}
(for an explicit expression see ref. [\cite{infrared,quadratic}]).
The quadratic form of the above equation obviously derives from the second
order formalism we have used.

  Alternatively, in more usual terms, one can look for the eigenvalue of the
mass operator or center of mass Hamiltonian
$ H_{\rm CM} \equiv M = M_0 + V $,
with $V$ defined by $ M_0V+VM_0+V^2=U $. Neglecting the term $ V^2 $,
the linear potential $V$ can be obtained from $U$ simply by the
kinematic replacement
$
\sqrt{ (w_1+w_2) (w_1^\prime +w_2^\prime)\over w_1w_2w_1^\prime w_2^\prime}
\to {1\over 2\sqrt{w_1 w_2 w_1^\prime w_2^\prime}}
$.
Such expression is particularly useful for a comparison with models based
on potential. In particular in the static limit $V$ reduces to the Cornell
potential
\begin{equation}
V = - {4 \over 3} {\alpha_{\rm s} \over r } + \sigma r \, ,
    \label{eq:static}
\end{equation}
in the semirelativistic limit (up to ${1 \over m^2}$ terms after
an appropriate Foldy-Wouthuysen transformation) equals the
potential discussed in ref. [\cite{barch}]. If the spin dependent
terms are neglected but full relativistic kinematics is kept, it
becomes identical to the potential of the (relativistic) flux tube
model \cite{bmp}.


\section{Running coupling constant and effective mass}

   In ref. [\cite{quadratic}]  the spectrum was evaluated for both the 
operators $M^2$ and $H_{\rm CM}$ introduced in the preceding section, omitting
the spin-orbit terms in the potential but including the hyperfine ones.
As we told, with fixed coupling constant and quark masses, a general good
fit of the data was obtained over the entire calculable spectrum with,
however, the relevant exception of the light pseudoscalar mesons. At the
light of the idea that such mesons should be Goldston massless particles
in the chiral limit, this is not surprising \cite{chiral}. The results
obtained in ref. [\cite{infrared}] suggest, however,
that the situation can be  greatly improved using a running
coupling constant and an effective light quark mass function of the
momentum, as implied in (\ref{eq:dshom}) for a kernel of the type
(\ref{eq:imom}).

   At one loop the running coupling constant in
QCD is usually written as
\begin{equation}
  \alpha_{\rm s} ( Q^{2} ) = \frac{ 4 \pi }{ \beta_{0}
  \ln{ ( Q^{2} / \Lambda^{2} ) } }
\label{eq:runcst}
\end{equation}
$ Q $ being the relevant energy scale,
$ \beta_{0} = 11 - \frac{2}{3}  N_{\rm f} $,
and $ N_{\rm f} $ the number of flavors
with masses smaller than $ Q $. Such expression
becomes singular and completely inadequate as
$ Q^{2} $ approaches $ \Lambda^{2} $.

\begin{figure}[htbp!]
    \leavevmode
    \setlength{\unitlength}{1.0mm}
    \begin{picture}(140,60)
      \put(27,7){\epsfxsize=2.5in\epsfbox{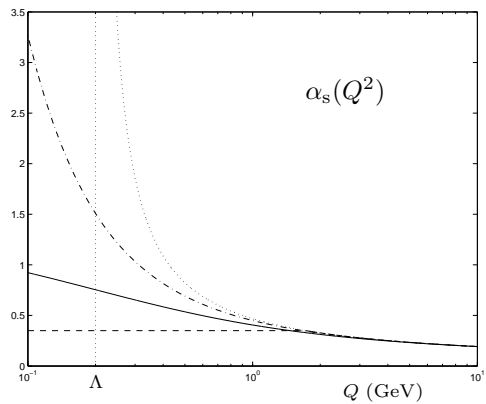}}
      \put(36.2,7){ $ {}_{\Lambda} $ }
      \put(65,45){ $ \alpha_{\rm s} (Q^{2}) $ }
      \put(70,6){ $ {}_{Q \; ({\rm GeV})} $ }
   \end{picture}
\vspace{-1cm}
\caption{Running coupling constant
$ \alpha_{\rm s} (Q^{2}) $
on logarithmic scale.
Perturbative expression (dotted line),
truncation prescription (dashed line),
Shirkov-Solovtsov prescription (full line),
Nesterenko prescription (dot-dashed line).}
\label{figalpha}
\end{figure}

The most naive modification of eq. (\ref{eq:runcst})
consists in saturating $\alpha_{\rm s} (Q^{2})$
to a certain maximum value $\bar{\alpha}_{\rm s}$ as
$Q^2$ decrease (fig. \ref{figalpha})
and in treating this value as a phenomenological parameter
(truncation prescription).

Alternatively
Shirkov and Solovtsov \cite{shirkov} replace
(\ref{eq:runcst}) with
\begin{equation}
  \alpha_{\rm s} ( Q^{2} ) = \frac{ 4 \pi }{
   \beta_{0} } \left(
  \frac{1}{ \ln{ ( Q^{2} / \Lambda^{2} ) } } +
  \frac{ \Lambda^{2} }{ \Lambda^{2} - Q^{2} } \right).
\label{eq:runshk}
\end{equation}
This remains regular for $ Q^{2} = \Lambda^{2} $ and
has a finite $ \Lambda $ independent limit,
$ \alpha_{\rm s}(0) = 4 \pi / \beta_{0} $,
for $ Q^{2} \rightarrow 0 $. Eq. (\ref{eq:runshk}) is
obtained assuming
a dispersion relation with a cut
from
$-\infty<Q^2<0$ and applying
(\ref{eq:runcst}) to
the evaluation of the
spectral function alone.

In the quark-antiquark bound state
problem the variable $ Q^{2} $
can be identified
with the squared momentum
transfer
$ {\bf Q}^{2} = ( {\bf k} - {\bf k}^{\prime} )^{2} \! $.
Then, typically,
$ \; \langle {\bf Q}^{2} \rangle \; $
ranges ~between~ $ \; ( 0.1 \, {\rm GeV} )^{2} $
and
$ ( 1 \, {\rm GeV} )^{2} $
for different quark masses
and internal excitations
and values
of
$ {\bf Q}^{2} $ smaller than
$ \Lambda^{2} $ can be important.
The specific infrared behavior is
therefore~ expected
to affect~ the spectrum
and other properties of mesons.

   Coming to the quark masses, notice that, even neglecting the spin
dependent part in $\hat H^{(2)}(k)$ (consistently with what we do for
$ \langle {\bf k} \vert U \vert {\bf k}^{\prime} \rangle $), we could more
sensibly write $ \hat H^{(2)}(k) = i/(k^2 - m^2 - \Gamma_0(k) )$
in the reduction procedure, rather
than simply identify such quantity  with its
free expression ($\Gamma_0(k)=0$).
However, since the $ I_a(Q,p,p') $ are not formally
covariant, but given in terms of C.M. variables, $ \Gamma_0(k)$ must
depend separately on
$ k_{0}^{2} $ and ${\bf k}^{2}$.
Then the pole of $ H_{2}(k) $, defined by
\begin{equation}
  k_0^2 - {\bf k}^2 - m^2 -
  \Gamma(k_0^2,{\bf k}^2) = 0 \,,
\end{equation}
could be written
\begin{equation}
  k_0^2 = m_{\rm eff}^2( \vert {\bf k} \vert ) + {\bf k}^2 \, ,
\end{equation}
where $ m_{\rm eff}^{2}( \vert {\bf k} \vert ) $ would be a
$ \vert {\bf k} \vert $ dependent expression that is expected to
approach the current $ m^{2} $ for
$ \vert {\bf k} \vert \rightarrow 0 $
and to increase toward a kind of constituent $ m^{\prime 2} $
as $ \vert {\bf k} \vert $ increases
\footnote{Take into account that from the virial theorem for a linear
potential in the extreme relativistic approximation we have
$ \langle \vert {\bf k} \vert \rangle = \sigma \langle r \rangle $
and a large $ \vert {\bf k} \vert $ corresponds to a peripheral
interaction or, what is the same, to a small $ \vert {\bf Q} \vert $.}.
Then, eventually, we obtain the same operator $ M^2 $ as given
above, but with $ m_{1}$ and $ m_{2} $ replaced by expressions like
$ m_{\rm eff}^{2}( \vert {\bf k} \vert ) $.

 As matter of fact we have not tried in this paper to evaluate $\Gamma(p)$
from first principles by solving eq. (\ref{eq:dshom}) but have simply
parametrized $ m_{\rm eff}^{2} ( \vert {\bf k} \vert ) $ as
a polynomial in $ \vert {\bf k} \vert $ for the light quarks and used
a fixed mass as in the preceding papers for strange and heavy quarks.



\section{Numerical procedure and results}

In the calculation of the eigenvalues of the operator $M^2$ introduced in
sect. 4 we have used the same approximations and the same numerical procedure
as in the previous papers. We have retained only the hyperfine part of the
complicate spin dependent terms (spin orbit and tensorial ones) occurring
in the potential $U$. We have solved the eigenvalue equation for $H_{\rm CM}$
with $V$ equal its static limit (\ref{eq:static}) by the Rayleigh-Ritz
method, using the three dimensional Harmonic oscillator basis and
diagonalizing a $30 \times 30$ matrix. Then we have evaluated
$\langle \phi_n | M^2 |\phi_n\rangle $
for the eigenfunction $ \phi_n $ obtained in the first step.

\begin{figure}[htbp!]
  \begin{center}
    \leavevmode
    \setlength{\unitlength}{1.0mm}
    \begin{picture}(140,133)
      \put(0.3,89){\epsfxsize=2.2in\epsfbox{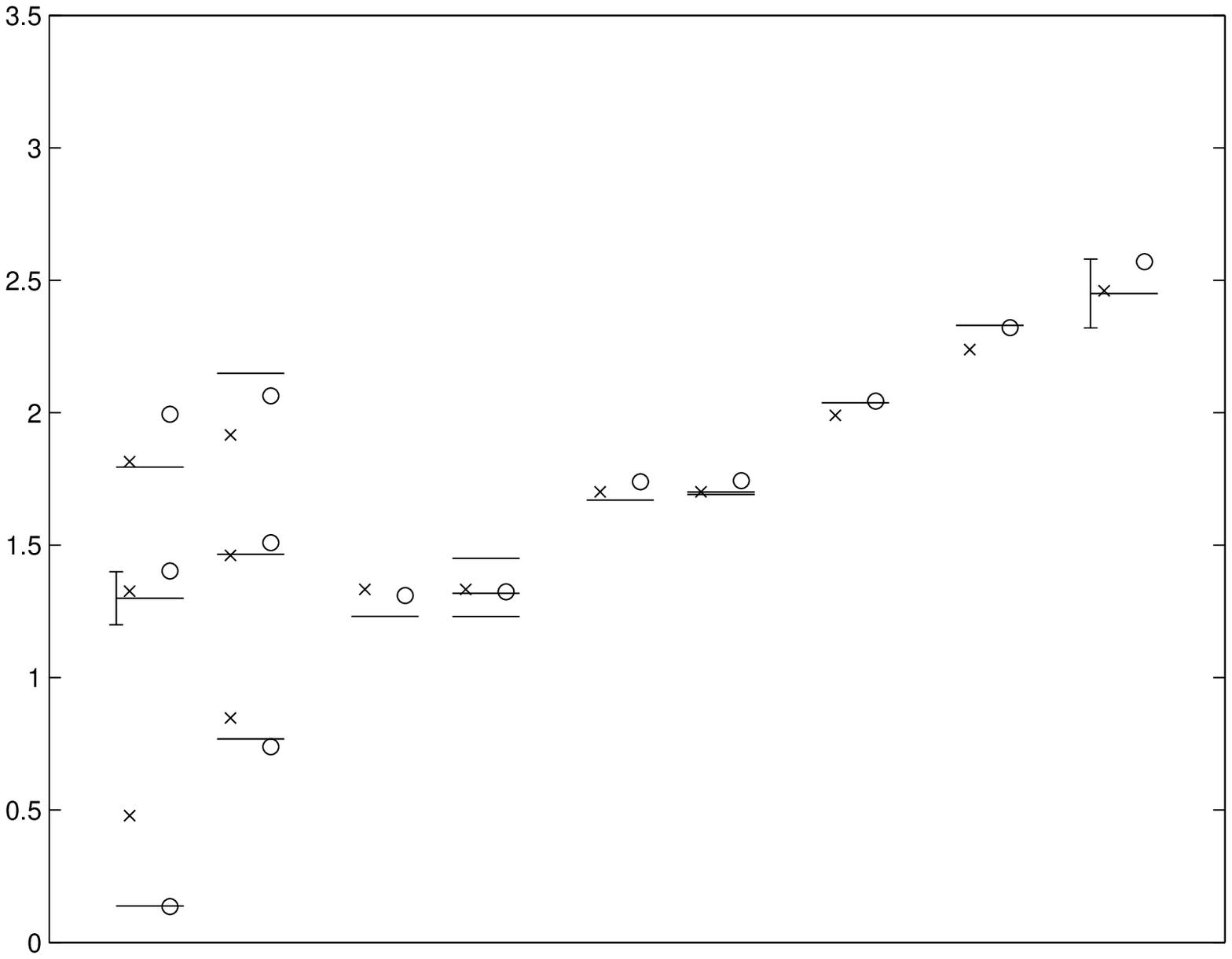}}
      \put(59.5,88.4){\epsfxsize=2.15in\epsfbox{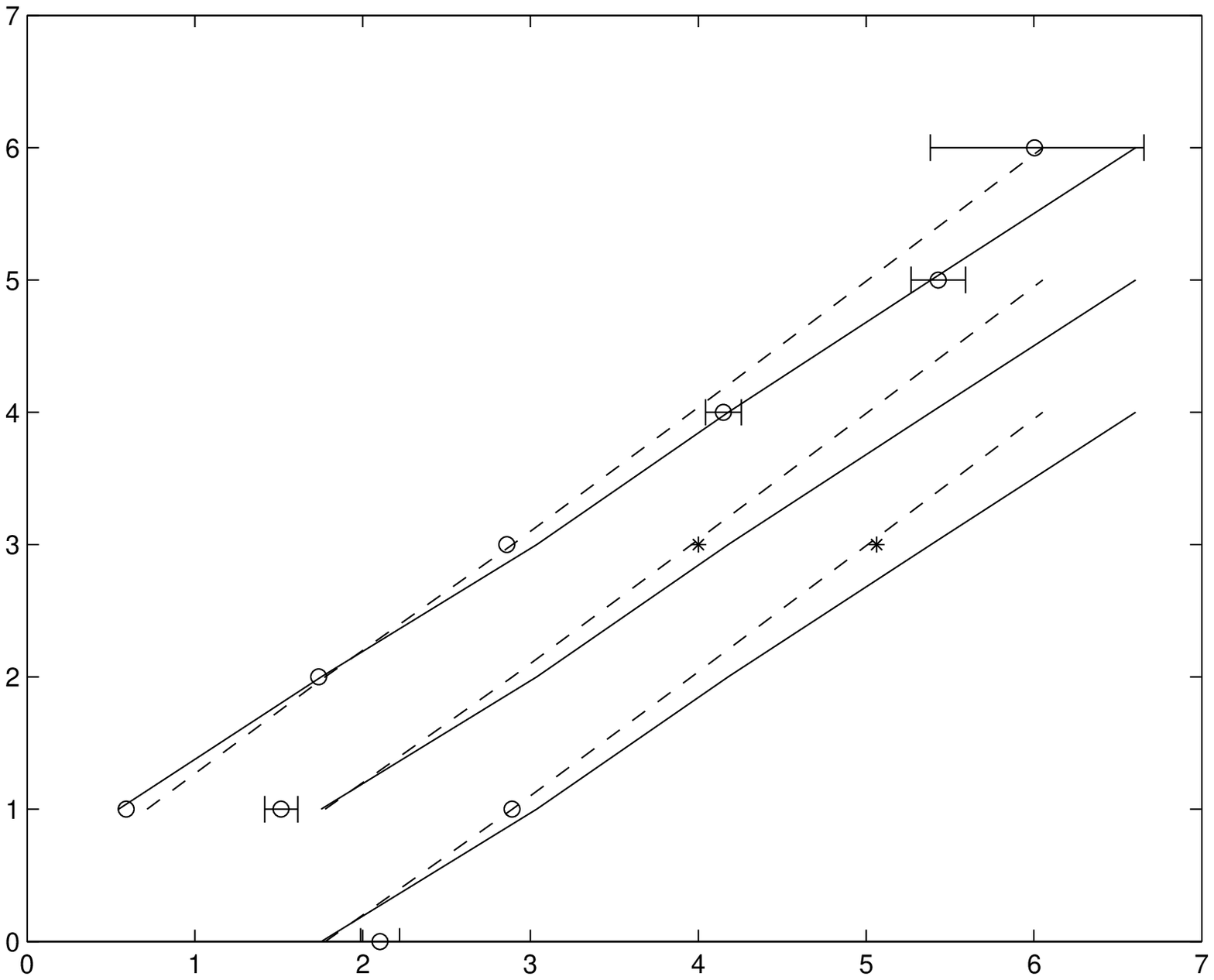}}
      \put(0.3,43){\epsfxsize=2.2in\epsfbox{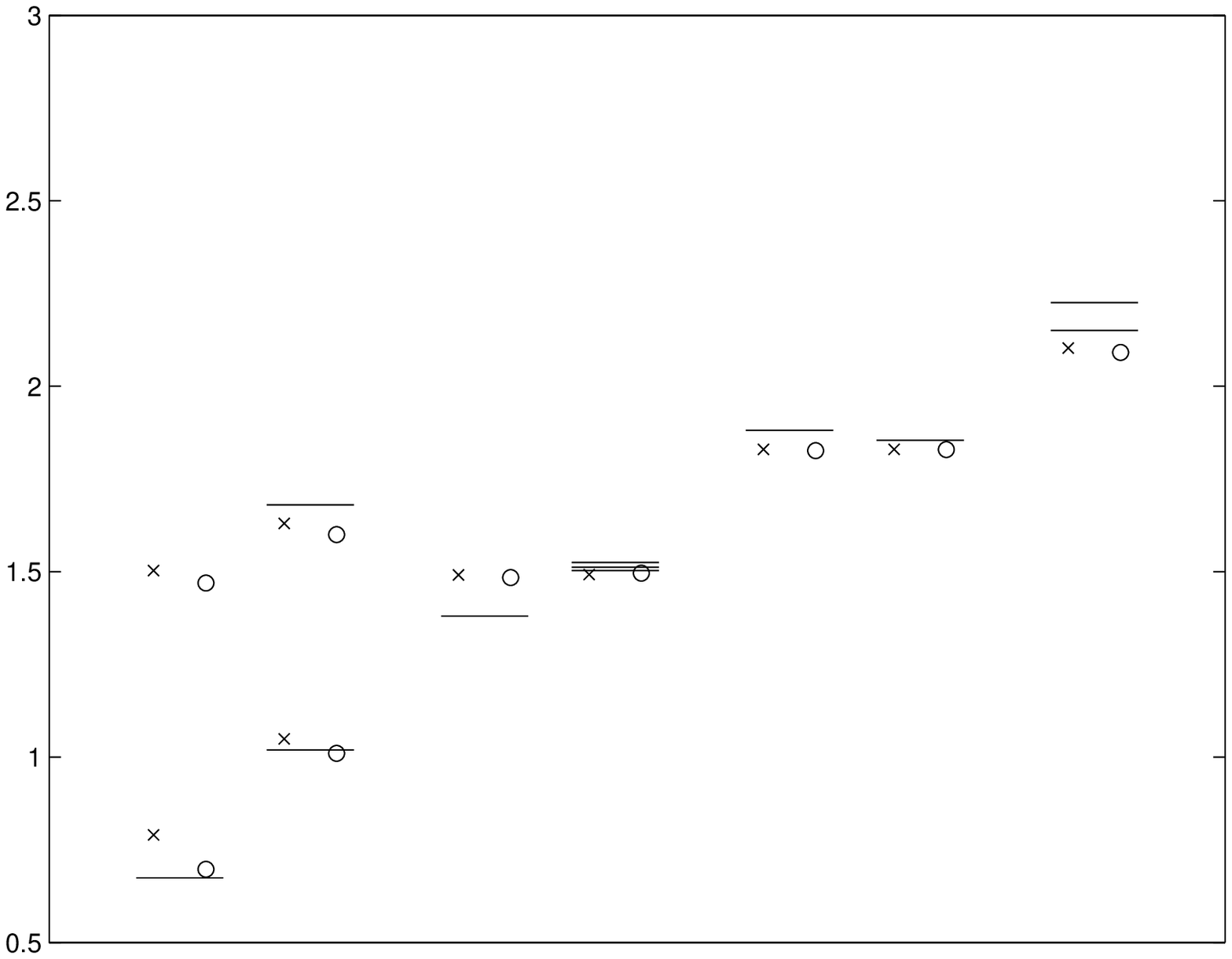}}
      \put(59.5,42.3){\epsfxsize=2.15in\epsfbox{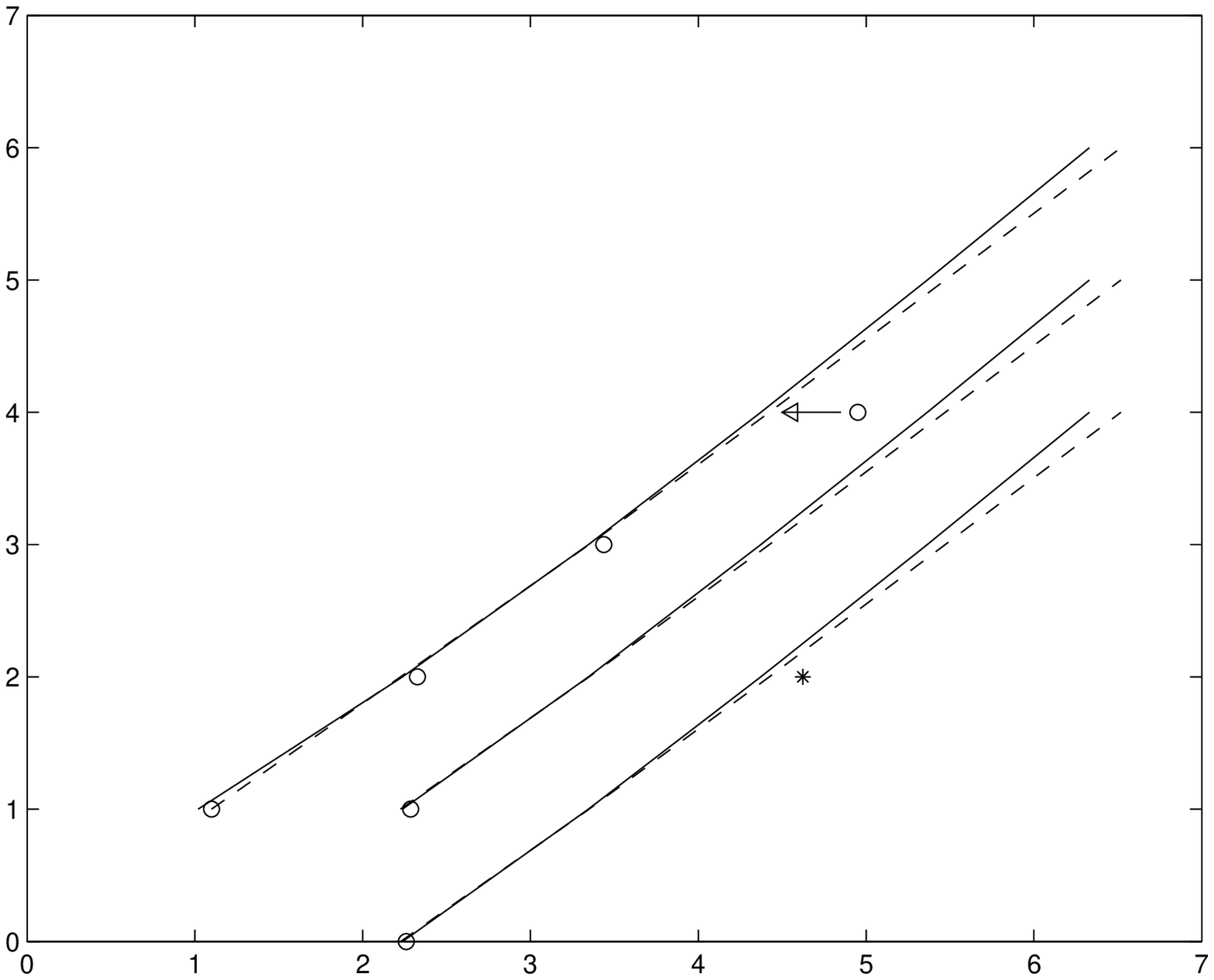}}
      \put(0.3,-3){\epsfxsize=2.2in\epsfbox{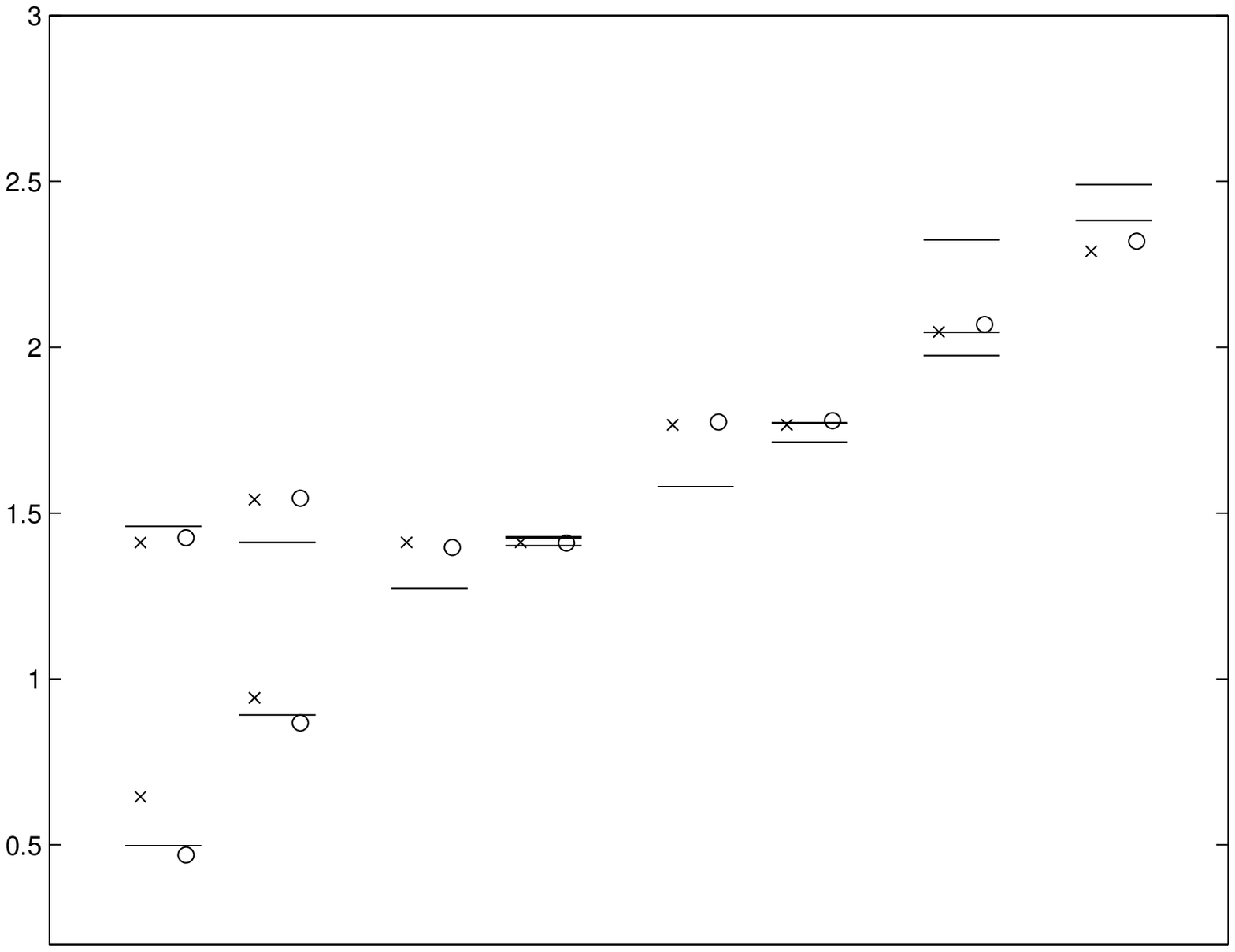}}
      \put(59.5,-4){\epsfxsize=2.15in\epsfbox{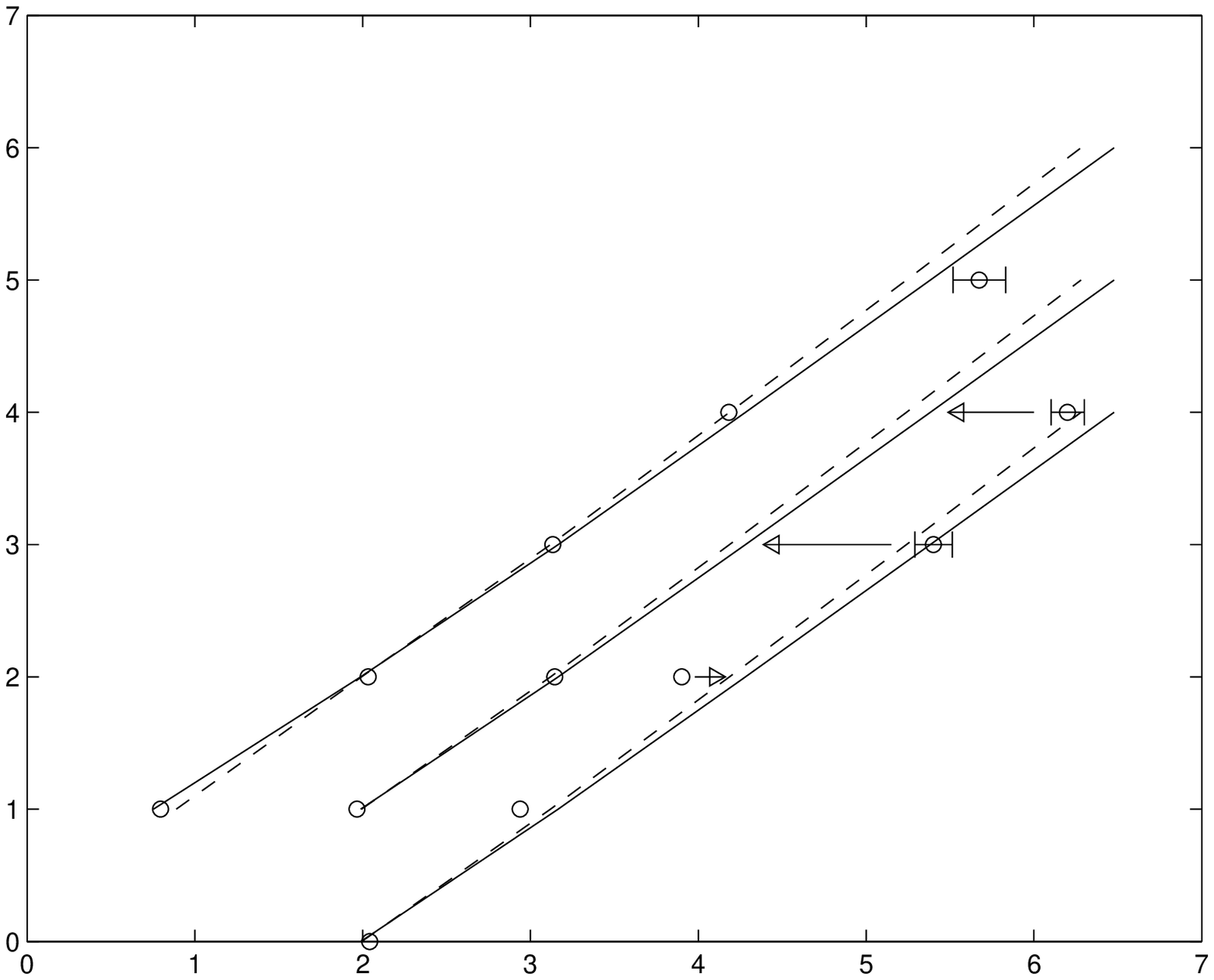}}
      \put(-3,135){ $ {}_{ ( {\rm GeV} ) } $ }
      \put(45,98){ $ q \bar{q} $ }
      \put(40,93){ $ q = u,d $ }
      \put(2,126){ $ {^{1} {\rm S}_{0}} $ }
      \put(7,126){ $ {^{3} {\rm S}_{1}} $ }
      \put(12,126){ $ {^{1} {\rm P}_{1}} $ }
      \put(17,126){ $ {^{3} {\rm P}_{J}} $ }
      \put(23,126){ $ {^{1} {\rm D}_{1}} $ }
      \put(28,126){ $ {^{3} {\rm D}_{J}} $ }
      \put(34.2,126){ $ {^{3} {\rm F}_{J}} $ }
      \put(40,126){ $ {^{3} {\rm G}_{J}} $ }
      \put(47,126){ $ {^{3} {\rm H}_{J}} $ }
      \put(70,126){ $ q \bar{q} $ }
      \put(62,121){ $ q = u,d $ }
      \put(88.7,107.15){ $ \ast $ }
      \put(96.8,107.15){ $ \ast $ }
      \put(60.5,127){ $ J $ }
      \put(106,91.5){ $ M^{2} $ }
      \put(106,126.7){ $ J = L + 1 $ }
      \put(106,120.7){ $ J = L $ }
      \put(106,114.7){ $ J = L - 1 $ }
      \put(56,98.5){ $ \rho (770) $ }
      \put(59.5,103.5){ $ a_{2} (1320) $ }
      \put(67.5,109){ $ \rho_{3} (1690) $ }
      \put(76.5,115){ $ a_{4} (2040) $ }
      \put(84.5,121){ $ \rho_{5} (2350) $ }
      \put(90,127.5){ $ a_{6} (2450) $ }
      \put(61.5,92.5){ $ a_{1} (1260) $ }
      \put(82,104){ $ X (2000) $ }
      \put(77,91){ $ a_{0} (1450) $ }
      \put(83.5,95){ $ \rho (1700) $ }
      \put(98.5,105.2){ $ \rho_{3} (2250) $ }
      \put(45,50){ $ s \bar{s} $ }
      \put(4.5,80){ $ {^{1} {\rm S}_{0}} $ }
      \put(10.5,80){ $ {^{3} {\rm S}_{1}} $ }
      \put(18,80){ $ {^{1} {\rm P}_{1}} $ }
      \put(24,80){ $ {^{3} {\rm P}_{J}} $ }
      \put(32,80){ $ {^{1} {\rm D}_{1}} $ }
      \put(38,80){ $ {^{3} {\rm D}_{J}} $ }
      \put(46.2,80){ $ {^{3} {\rm F}_{J}} $ }
      \put(10.5,46.7){ $ \eta_{s} $ }
      \put(70,80){ $ s \bar{s} $ }
      \put(93.4,55.1){ $ \ast $ }
      \put(60.5,81){ $ J $ }
      \put(106,45.5){ $ M^{2} $ }
      \put(106,80.5){ $ J = L + 1 $ }
      \put(106,74.5){ $ J = L $ }
      \put(106,68.5){ $ J = L - 1 $ }
      \put(56,52){ $ \phi (1020) $ }
      \put(63,57){ $ f_{2}^{\prime} (1525) $ }
      \put(71.2,63){ $ \phi_{3} (1850) $ }
      \put(78.7,69){ $ f_{J} (2220) $ }
      \put(65.7,46.8){ $ f_{1} (1510) $ }
      \put(80,45){ $ f_{0} (1500) $ }
      \put(95.2,54){ $ f_{2} (2150) $ }
      \put(45,6){ $ q \bar{s} $ }
      \put(40,1){ $ q = u,d $ }
      \put(2,34){ $ {^{1} {\rm S}_{0}} $ }
      \put(7,34){ $ {^{3} {\rm S}_{1}} $ }
      \put(14,34){ $ {^{1} {\rm P}_{1}} $ }
      \put(19,34){ $ {^{3} {\rm P}_{J}} $ }
      \put(26,34){ $ {^{1} {\rm D}_{1}} $ }
      \put(31,34){ $ {^{3} {\rm D}_{J}} $ }
      \put(39,34){ $ {^{3} {\rm F}_{J}} $ }
      \put(46,34){ $ {^{3} {\rm G}_{J}} $ }
      \put(70,34){ $ q \bar{s} $ }
      \put(62,29){ $ q = u,d $ }
      \put(60.5,35){ $ J $ }
      \put(106,-1){ $ M^{2} $ }
      \put(106,34.5){ $ J = L + 1 $ }
      \put(106,28.5){ $ J = L $ }
      \put(106,22.5){ $ J = L - 1 $ }
      \put(54.5,6){ $ K^{*} (892) $ }
      \put(60,11){ $ K_{2}^{*} (1430) $ }
      \put(68,16.5){ $ K_{3}^{*} (1780) $ }
      \put(76,22.5){ $ K_{4}^{*} (2045) $ }
      \put(84.5,28.5){ $ K_{5}^{*} (2380) $ }
      \put(63,0.5){ $ K_{1} (1400) $ }
      \put(75,6.2){ $ K_{2} (1770) $ }
      \put(79.5,-1){ $ K_{0}^{*} (1430) $ }
      \put(84.5,2.4){ $ K^{*} (1680) $ }
      \put(91.5,7){ $ K_{2}^{*} (1980) $ }
      \put(98,12){ $ K_{3} (2320) $ }
      \put(107,18){ $ K_{4} (2500) $ }
   \end{picture}
  \end{center}
\caption{Light-light quarkonium spectrum and Regge trajectories.
$ n_{f} = 4 $, $ \Lambda = 0.2 \; {\rm GeV} $.
Crosses and dashed line:
truncation prescription, $ \bar\alpha_{\rm s} = 0.35 $,
$ \sigma = 0.2 \; {\rm GeV}^{2} $,
$ m_{s} = 0.2 \; {\rm GeV}  $,
$ m_{u} = m_{d} = 0.01 \; {\rm GeV}  $.
Circlets and full line:
Shirkov-Solovtsov
$ \alpha_{\rm s}( Q^{2} ) $,
$ \sigma = 0.18 \; {\rm GeV}^{2} $,
$ m_{s} = 0.381 \; {\rm GeV}  $,
$ m_{u}^{2} = m_{d}^{2} = 0.121 \, k - 0.025 \, k^{2}
              + 0.25 \, k^{4}  $.}
\label{figuussus}
\end{figure}

\begin{figure}[htbp!]
  \begin{center}
    \leavevmode
    \setlength{\unitlength}{1.0mm}
    \begin{picture}(140,85)
      \put(0.3,42){\epsfxsize=2.2in\epsfbox{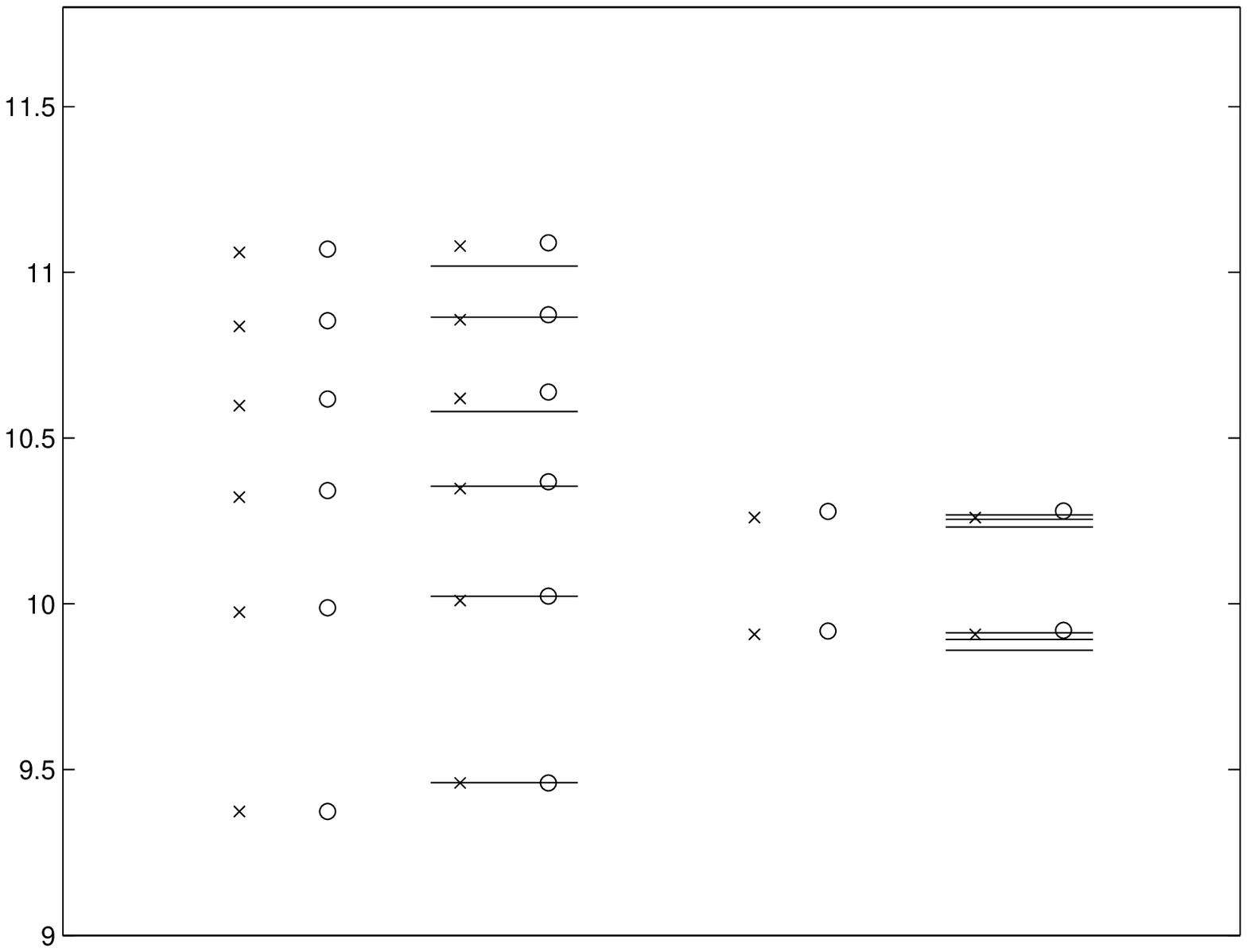}}
      \put(58.5,41.8){\epsfxsize=2.2in\epsfbox{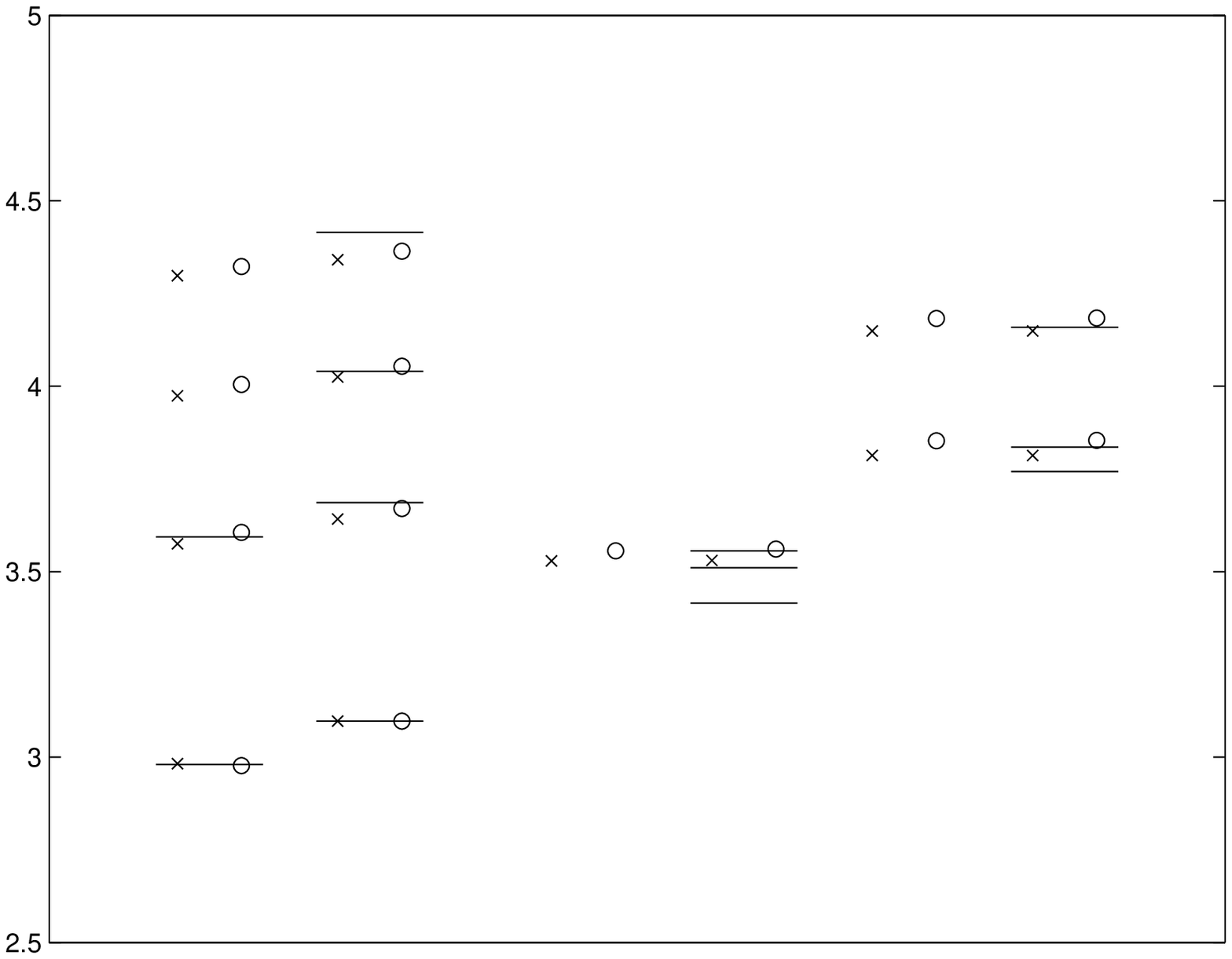}}
      \put(1,-3){\epsfxsize=2.2in\epsfbox{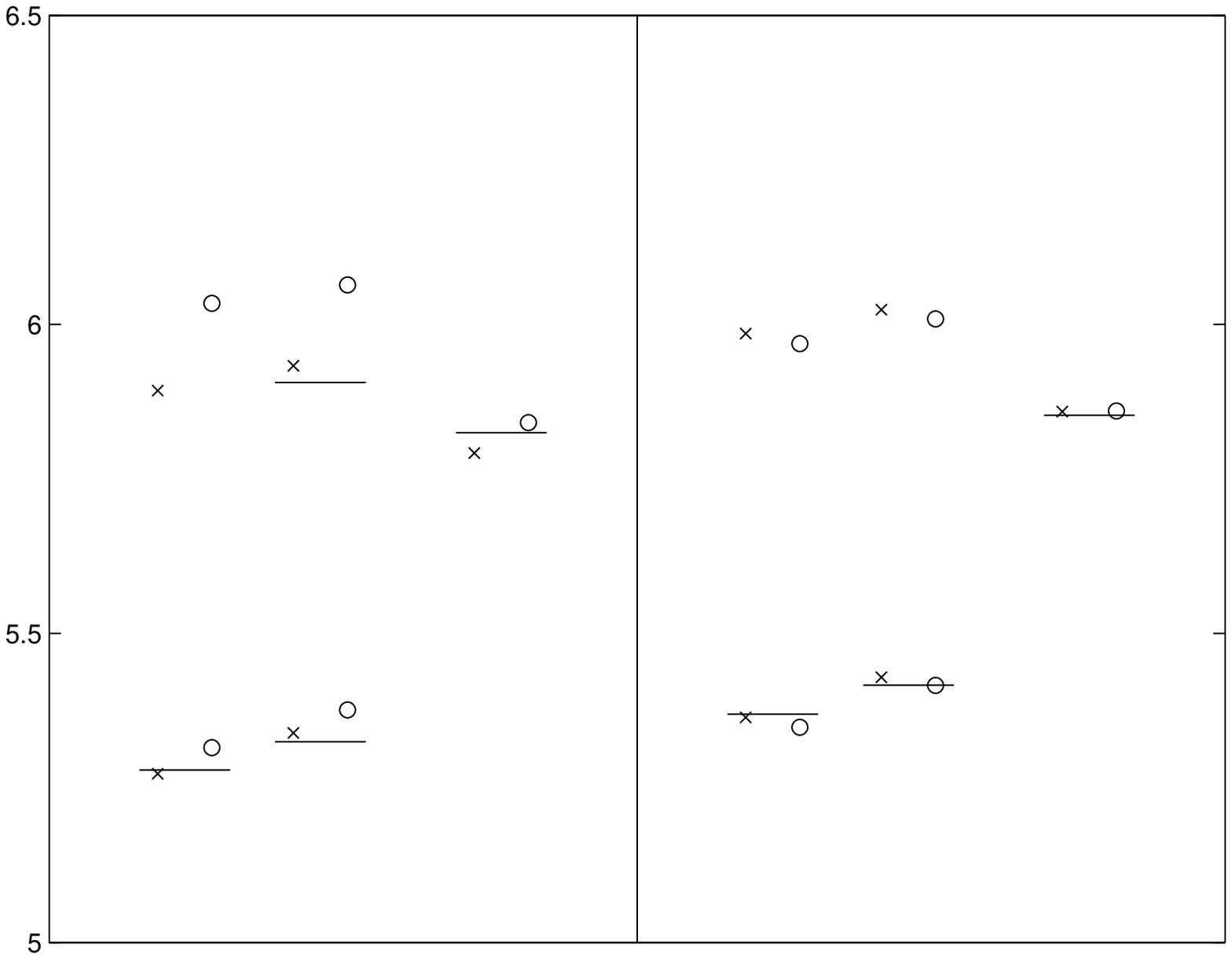}}
      \put(58.5,-2.8){\epsfxsize=2.2in\epsfbox{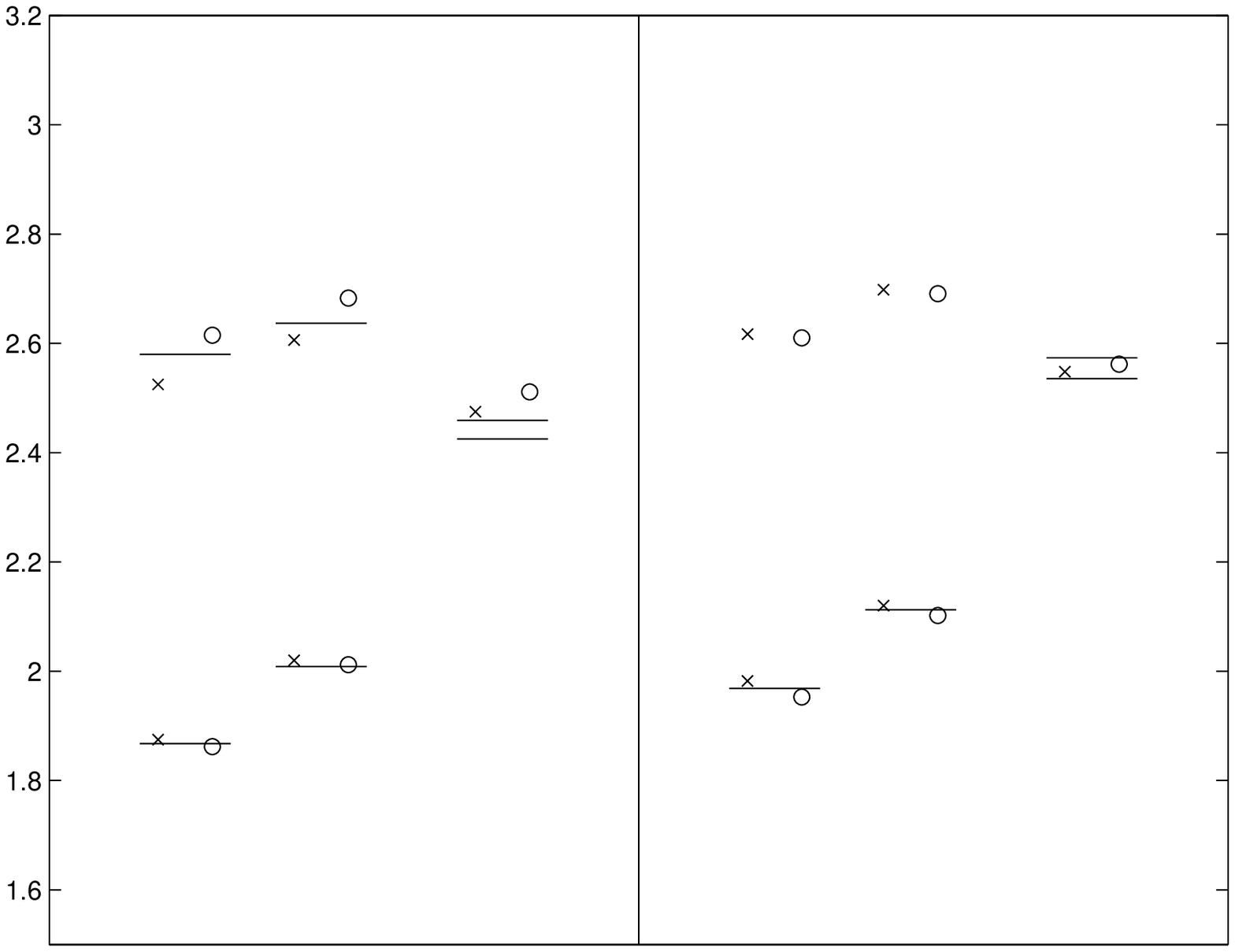}}
      \put(45,48){ $ b \bar{b} $ }
      \put(-2,87){ $ {}_{ ( {\rm GeV} ) } $ }
      \put(9,79){ $ {^{1} {\rm S}_{0}} $ }
      \put(19,79){ $ {^{3} {\rm S}_{1}} $ }
      \put(32,79){ $ {^{1} {\rm P}_{1}} $ }
      \put(42,79){ $ {^{3} {\rm P}_{J}} $ }
      \put(15,76){ n }
      \put(15.5,73){ $ {}_{6} $ }
      \put(15.5,70){ $ {}_{5} $ }
      \put(15.5,66){ $ {}_{4} $ }
      \put(15.5,62.5){ $ {}_{3} $ }
      \put(15.5,57){ $ {}_{2} $ }
      \put(15.5,49){ $ {}_{1} $ }
      \put(37.5,76){ n }
      \put(38,61){ $ {}_{2} $ }
      \put(38,56){ $ {}_{1} $ }
      \put(49.5,74){ $ {}_{J} $ }
      \put(51.5,62){ $ {}_{2} $ }
      \put(50,61){ $ {}_{1} $ }
      \put(48.5,60){ $ {}_{0} $ }
      \put(51.5,57){ $ {}_{2} $ }
      \put(50,56){ $ {}_{1} $ }
      \put(48.5,55){ $ {}_{0} $ }
      \put(103,48){ $ c \bar{c} $ }
      \put(56,87){ $ {}_{ ( {\rm GeV} ) } $ }
      \put(64,79){ $ {^{1} {\rm S}_{0}} $ }
      \put(71,79){ $ {^{3} {\rm S}_{1}} $ }
      \put(79.5,79){ $ {^{1} {\rm P}_{1}} $ }
      \put(86.5,79){ $ {^{3} {\rm P}_{J}} $ }
      \put(95,79){ $ {^{1} {\rm D}_{1}} $ }
      \put(102,79){ $ {^{3} {\rm D}_{J}} $ }
      \put(69,76){ n }
      \put(69.5,73){ $ {}_{4} $ }
      \put(69.5,67.5){ $ {}_{3} $ }
      \put(69.5,61){ $ {}_{2} $ }
      \put(69.5,51){ $ {}_{1} $ }
      \put(85.5,76){ n }
      \put(86,59){ $ {}_{1} $ }
      \put(100.5,76){ n }
      \put(101,70){ $ {}_{2} $ }
      \put(101,64){ $ {}_{1} $ }
      \put(93.5,74){ $ {}_{J} $ }
      \put(94,61){ $ {}_{2} $ }
      \put(94,59){ $ {}_{1} $ }
      \put(94,57){ $ {}_{0} $ }
      \put(108.5,74){ $ {}_{J} $ }
      \put(109,70){ $ {}_{1} $ }
      \put(109,65){ $ {}_{2} $ }
      \put(109,63){ $ {}_{1} $ }
      \put(21,4){ $ q \bar{b} $ }
      \put(15,0){ $ q = u,d $ }
      \put(5,34){ $ {^{1} {\rm S}_{0}} $ }
      \put(12,34){ $ {^{3} {\rm S}_{1}} $ }
      \put(21,34){ $ {\rm P} $ }
      \put(10,30){ n }
      \put(10.5,24){ $ {}_{2} $ }
      \put(10.5,6){ $ {}_{1} $ }
      \put(18,30){ n }
      \put(18.5,20){ $ {}_{1} $ }
      \put(47,2){ $ s \bar{b} $ }
      \put(32,34){ $ {^{1} {\rm S}_{0}} $ }
      \put(38,34){ $ {^{3} {\rm S}_{1}} $ }
      \put(47.5,34){ $ {\rm P} $ }
      \put(36.5,30){ n }
      \put(37,25){ $ {}_{2} $ }
      \put(37,8.5){ $ {}_{1} $ }
      \put(44.5,30){ n }
      \put(45,21){ $ {}_{1} $ }
      \put(79,4){ $ q \bar{c} $ }
      \put(73,0){ $ q = u,d $ }
      \put(64,34){ $ {^{1} {\rm S}_{0}} $ }
      \put(70,34){ $ {^{3} {\rm S}_{1}} $ }
      \put(79.5,34){ $ {\rm P} $ }
      \put(67.5,30){ n }
      \put(68,24){ $ {}_{2} $ }
      \put(68,8){ $ {}_{1} $ }
      \put(75.5,30){ n }
      \put(76,20.5){ $ {}_{1} $ }
      \put(105,2){ $ s \bar{c} $ }
      \put(90,34){ $ {^{1} {\rm S}_{0}} $ }
      \put(96,34){ $ {^{3} {\rm S}_{1}} $ }
      \put(105.5,34){ $ {\rm P} $ }
      \put(94.2,30){ n }
      \put(94.7,25.5){ $ {}_{2} $ }
      \put(94.7,10.5){ $ {}_{1} $ }
      \put(102.2,30){ n }
      \put(102.7,23.1){ $ {}_{1} $ }
    \end{picture}
  \end{center}
\caption{Heavy-heavy and heavy-light quarkonium spectrum.
$ n_{f} = 4 $, $ \Lambda = 0.2 \; {\rm GeV} $.
Crosses:
truncation prescription,
$ \bar\alpha_{\rm s} = 0.35 $,
$ \sigma = 0.2 \; {\rm GeV}^{2} $,
$ m_{b} = 4.763 \; {\rm GeV}  $,
$ m_{c} = 1.394 \; {\rm GeV}  $,
$ m_{s} = 0.2 \; {\rm GeV}  $,
$ m_{u} = m_{d} = 0.01 \; {\rm GeV}  $.
Circlets:
Shirkov-Solovtsov
$ \alpha_{\rm s}( Q^{2} ) $,
$ \sigma = 0.18 \; {\rm GeV}^{2} $,
$ m_{b} = 4.898 $,
$ m_{c} = 1.545 $,
$ m_{s} = 0.381 \; {\rm GeV} $,
$ m_{u}^{2} = m_{d}^{2} = 0.121 \, k - 0.025 \, k^{2}
              + 0.25 \, k^{4}  $.}
\label{figppelp}
\end{figure}

  In figs. \ref{figuussus} and \ref{figppelp}
we have represented graphically the results already obtained in ref.
[\cite{quadratic}] with a truncated running coupling constant (crosses)
and the new ones performed with the Shirkov-Solovtsov prescription
(circlets),
contrasted with the data (small lines) [\cite{data}].
In tables I--VII we give the corresponding numerical values.


  In both cases we have taken $N_{\rm f}=4$ and
$\Lambda=200\, {\rm MeV}$. With the truncation prescription, the light
quark masses were fixed on typical current values,
$ m_{u} = m_{d} = 10 \, {\rm MeV} $,
$ m_{s} = 200 \, {\rm MeV}$,
the maximum value for $\alpha_{\rm s}(Q^2)$, the
string tension and the heavy quark masses were treated as adjustable
parameters and chosen as $\bar\alpha_{\rm s}=0.35$,
$\sigma = 0.2\,{\rm GeV}^2$
(in order to reproduce the correct $J/\Psi - \eta_c$ separation and Regge
trajectory slope) and $m_c=1.394\, {\rm GeV}$, $m_b=4.763\, {\rm GeV}$
(in order to obtain exact masses for $J/\Psi$ and $\Upsilon$). On the
contrary, with the
Shirkov-Solovtsov prescription, we have fixed the string tension on
the commonly accepted value  $\sigma = 0.18\,{\rm GeV}^2$ and used
all masses as adjustable parameters. For the light quarks we have
taken (in $ {\rm GeV}^{2} $)

\begin{equation}
    m_u^2 = m_d^2 = 0.121 \, k - 0.025 \, k^2 + 0.25 \, k^4 \, ,
\label{eq:lqm}
\end{equation}
\noindent
($k$ being the quark momentum in the meson
C. M. frame), for the strange and the heavy quarks
$m_s=0.381 \, {\rm GeV} $, $m_c=1.545 \, {\rm GeV}$
and $m_b=4.898 \, {\rm GeV}$.

\begin{table}
\centering
{Table I. $ q \bar{q} \; (q = u,d) $,
$ n_{f} = 4 $, $ \Lambda = 0.2 \; {\rm GeV} $.
(a)
truncation prescription 
$ \bar\alpha_{\rm s} = 0.35 $,
$ \sigma = 0.2 \; {\rm GeV}^{2} $,
$ m_{u} = m_{d} = 0.01 \; {\rm GeV}  $. \\
(b)
Shirkov-Solovtsov
$ \alpha_{\rm s}( Q^{2} ) $,
$ \sigma = 0.18 \; {\rm GeV}^{2} $,
$ m_{u}^{2} = m_{d}^{2} = 0.121 \, k - 0.025 \, k^{2} +
0.25 \, k^{4} $.
\vspace{1mm}}
\begin{tabular}{ccccc}
\hline
 States &  & experimental values & (a) & (b)  \\
  &  & (MeV) & (MeV) & (MeV)  \\
\hline
$ 1 \, {^{1} {\rm S}_{0}} $ &
$
\left\{
\begin{array}{c}
\pi^{0}   \\
\pi^{\pm}
\end{array}
\right.
$
&
$
\left.
\begin{array}{c}
134.9764 \pm 0.0006  \\
139.56995 \pm 0.00035
\end{array}
\right\}
$ & 479 & 135  \\
$ 1 \, {^{3} {\rm S}_{1}} $ & $ \rho (770) $
& 768.5 $ \pm $ 0.6 & 846 & 739  \\
$ 1 \Delta {\rm SS} $ &  & 630 & 367 & 604  \\
$ 2 \, {^{1} {\rm S}_{0}} $ & $ \pi (1300) $
& 1300 $ \pm $ 100 & 1326 & 1402  \\
$ 2 \, {^{3} {\rm S}_{1}} $ & $ \rho (1450) $
& 1465 $ \pm $ 25 & 1461 & 1509  \\
$ 2 \Delta {\rm SS} $ &  & 165 & 135 & 107  \\
$ 3 \, {^{1} {\rm S}_{0}} $ & $ \pi (1800) $
& 1795 $ \pm $ 10 & 1815 & 1994  \\
$ 3 \, {^{3} {\rm S}_{1}} $ & $ \rho (2150) $
& 2149 $ \pm $ 17 & 1916 & 2064  \\
$ 3 \Delta {\rm SS} $ &  & 354 & 101 & 70  \\
\hline
$ 1 \, {^{1} {\rm P}_{1}} $ & $ b_{1} (1235) $
& 1231 $ \pm $ 10 & 1333 & 1310  \\
$
\begin{array}{c}
1 \, {^{3} {\rm P}_{2}} \\
1 \, {^{3} {\rm P}_{1}} \\
1 \, {^{3} {\rm P}_{0}}
\end{array}
$ &
$
\begin{array}{c}
a_{2} (1320) \\
a_{1} (1260) \\
a_{0} (1450)
\end{array}
$ &
$
\left.
\begin{array}{c}
1318.1 \pm 0.7 \\
1230 \pm 40 \\
1450 \pm 40
\end{array}
\right\}
1303
$ & 1333 & 1324  \\
\hline
$ 1 \, {^{1} {\rm D}_{2}} $ &
$ \pi_{2} (1670) $ & $ 1670 \pm 20 $ & 1701 & 1739  \\
$
\begin{array}{c}
1 \, {^{3} {\rm D}_{3}} \\
1 \, {^{3} {\rm D}_{2}} \\
1 \, {^{3} {\rm D}_{1}}
\end{array}
$ &
$
\begin{array}{c}
\rho_{3} (1690) \\    \\  \rho (1700)
\end{array}
$ &
$
\left.
\begin{array}{c}
1691 \pm 5  \\    \\ 1700 \pm 20
\end{array}
\right\}
$ & 1701 & 1743  \\
\hline
$ 1 \, {^{1} {\rm F}_{3}} $ & &  & 1990 & 2043  \\
$
\begin{array}{c}
1 \, {^{3} {\rm F}_{4}} \\
1 \, {^{3} {\rm F}_{3}} \\
1 \, {^{3} {\rm F}_{2}}
\end{array}
$ &
$
\begin{array}{c}
a_{4} (2040) \\ X(2000) \\
\\
\end{array}
$ &
$
\left.
\begin{array}{c}
2037 \pm 26  \\    \\   \\
\end{array}
\right\}
$ & 1990 & 2044  \\
\hline
$ 1 \, {^{1} {\rm G}_{4}} $ & & & 2238 & 2318  \\
$
\begin{array}{c}
1 \, {^{3} {\rm G}_{5}} \\
1 \, {^{3} {\rm G}_{4}} \\
1 \, {^{3} {\rm G}_{3}}
\end{array}
$ &
$
\begin{array}{c}
\rho_{5} (2350) \\  \\ \rho_{3} (2250)
\end{array}
$ &
$
\left.
\begin{array}{c}
2330 \pm 35  \\    \\   \\
\end{array}
\right\}
$ & 2238 & 2321  \\
\hline
$ 1 \, {^{1} {\rm H}_{5}} $ & & & 2460 & 2570  \\
$
\begin{array}{c}
1 \, {^{3} {\rm H}_{6}} \\
1 \, {^{3} {\rm H}_{5}} \\
1 \, {^{3} {\rm H}_{4}}
\end{array}
$ &
$
\begin{array}{c}
a_{6} (2450) \\  \\   \\
\end{array}
$ &
$
\left.
\begin{array}{c}
2450 \pm 130  \\    \\   \\
\end{array}
\right\}
$ & 2460 & 2570 \\
\hline
\end{tabular}
\end{table}
\begin{table}
\centering
{Table II. $ s \bar{s} $,
$ n_{f} = 4 $, $ \Lambda = 0.2 \; {\rm GeV} $.
(a)
truncation prescription $ \bar\alpha_{\rm s} = 0.35 $,
$ \sigma = 0.2 \; {\rm GeV}^{2} $,
$ m_{s} = 0.2 \; {\rm GeV} $.
(b)
Shirkov-Solovtsov
$ \alpha_{\rm s}( Q^{2} ) $,
$ \sigma = 0.18 \; {\rm GeV}^{2} $,
$ m_{s} = 0.381 \; {\rm GeV} $.
\vspace{1mm}}
\begin{tabular}{ccccc}
\hline
 States &  & experimental values & (a) & (b)  \\
  &  & (MeV) & (MeV) & (MeV)  \\
\hline
$ 1 \, {^{1} {\rm S}_{0}} $ &
$ \eta_{\rm s} $ &
$
\left\{
\begin{array}{c}
\eta (547) \\
\eta^{\prime} (958)
\end{array}
\right\}
674
$
& 790 & 697  \\
$ 1 \, {^{3} {\rm S}_{1}} $ & $ \phi (1020) $
& 1019.413 $ \pm $ 0.008 & 1049 & 1010  \\
$ 1 \Delta {\rm SS} $ &  & 335.3 $ \pm $ 0.1 & 259 & 313  \\
$ 2 \, {^{1} {\rm S}_{0}} $ & & & 1503 & 1469  \\
$ 2 \, {^{3} {\rm S}_{1}} $ & $ \phi (1680) $
& 1680 $ \pm $ 20 & 1630 & 1600  \\
$ 2 \Delta {\rm SS} $ &  &  & 127 & 131  \\
$ 3 \, {^{1} {\rm S}_{0}} $ & & & 1977 & 1934  \\
$ 3 \, {^{3} {\rm S}_{1}} $ & & & 2068 & 2023  \\
$ 3 \Delta {\rm SS} $ &  &  & 91 & 89  \\
\hline
$ 1 \, {^{1} {\rm P}_{1}} $ & $ h_{1} (1380) $
& 1380 $ \pm $ 20 & 1491 & 1484  \\
$
\begin{array}{c}
1 \, {^{3} {\rm P}_{2}} \\
1 \, {^{3} {\rm P}_{1}} \\
1 \, {^{3} {\rm P}_{0}}
\end{array}
$ &
$
\begin{array}{c}
f_{2}^{\prime} (1525) \\
f_{1} (1510) \\
f_{0} (1500)
\end{array}
$ &
$
\left.
\begin{array}{c}
1525 \pm 5 \\
1512 \pm 4 \\
1503 \pm 11
\end{array}
\right\}
1518
$ & 1492 & 1496  \\
\hline
$ 1 \, {^{1} {\rm D}_{2}} $ &
$ \eta_{2} (1870) $ & $ 1881 \pm 32 \pm 40 $ & 1830 & 1826  \\
$
\begin{array}{c}
1 \, {^{3} {\rm D}_{3}} \\
1 \, {^{3} {\rm D}_{2}} \\
1 \, {^{3} {\rm D}_{1}}
\end{array}
$ &
$
\begin{array}{c}
\phi_{3} (1850) \\    \\  \\
\end{array}
$ &
$
\left.
\begin{array}{c}
1854 \pm 7  \\    \\  \\
\end{array}
\right\}
$ & 1830 & 1829  \\
\hline
$ 1 \, {^{1} {\rm F}_{3}} $ & & & 2103 & 2089 \\
$
\begin{array}{c}
1 \, {^{3} {\rm F}_{4}} \\
1 \, {^{3} {\rm F}_{3}} \\
1 \, {^{3} {\rm F}_{2}}
\end{array}
$ &
$
\begin{array}{c}
f_{J} (2220) \\  \\ f_{2} (2150)  \\
\end{array}
$ &
$
\left.
\begin{array}{c}
2225 \pm 6  \\   \\   \\
\end{array}
\right\}
$ & 2103 & 2091  \\
\hline
$ 1 \, {^{1} {\rm G}_{4}} $ & & & 2339 & 2313  \\
$
\begin{array}{c}
1 \, {^{3} {\rm G}_{5}} \\
1 \, {^{3} {\rm G}_{4}} \\
1 \, {^{3} {\rm G}_{3}}
\end{array}
$ &
$
\begin{array}{c}
   \\  \\   \\
\end{array}
$ &
$
\left.
\begin{array}{c}
  \\    \\   \\
\end{array}
\right\}
$ & 2339 & 2316  \\
\hline
$ 1 \, {^{1} {\rm H}_{5}} $ & & & 2553 & 2516  \\
$
\begin{array}{c}
1 \, {^{3} {\rm H}_{6}} \\
1 \, {^{3} {\rm H}_{5}} \\
1 \, {^{3} {\rm H}_{4}}
\end{array}
$ &
$
\begin{array}{c}
   \\  \\   \\
\end{array}
$ &
$
\left.
\begin{array}{c}
    \\    \\   \\
\end{array}
\right\}
$ & 2553 & 2516  \\
\hline
\end{tabular}
\end{table}
\begin{table}
\centering
{Table III. $ q \bar{s} \; (q = u,d) $,
$ n_{f} = 4 $, $ \Lambda = 0.2 \; {\rm GeV} $.
(a)
truncation prescription
$ \bar\alpha_{\rm s} = 0.35 $,
$ \sigma = 0.2 \; {\rm GeV}^{2} $,
$ m_{s} = 0.2 \; {\rm GeV} $,
$ m_{u} = m_{d} = 0.01 \; {\rm GeV}  $. \\
(b)
Shirkov-Solovtsov
$ \alpha_{\rm s}( Q^{2} ) $,
$ \sigma = 0.18 \; {\rm GeV}^{2} $,
$ m_{s} = 0.381 \; {\rm GeV} $,
$ m_{u}^{2} = m_{d}^{2} = 0.121 \, k - 0.025 \, k^{2} +
0.25 \, k^{4} $.
\vspace{1mm}}
\begin{tabular}{ccccc}
\hline
 States &  & experimental values & (a) & (b)  \\
  &  & (MeV) & (MeV) & (MeV)  \\
\hline
$ 1 \, {^{1} {\rm S}_{0}} $ &
$
\left\{
\begin{array}{c}
K^{0}   \\
K^{\pm}
\end{array}
\right.
$
&
$
\left.
\begin{array}{c}
497.672 \pm 0.031  \\
493.677 \pm 0.016
\end{array}
\right\}
$ & 645 & 470  \\
$ 1 \, {^{3} {\rm S}_{1}} $ & $ K^{\ast} (892) $
& 891.59 $ \pm $ 0.24 & 943 & 868  \\
$ 1 \Delta {\rm SS} $ &  & 393.9 $ \pm $ 0.3 & 298 & 398  \\
$ 2 \, {^{1} {\rm S}_{0}} $ & $ K(1460) $ & & 1411 & 1426  \\
$ 2 \, {^{3} {\rm S}_{1}} $ & $ K^{\ast} (1410) $
& 1412 $ \pm $ 12 & 1541 & 1545  \\
$ 2 \Delta {\rm SS} $ &  &  & 130 & 119  \\
$ 3 \, {^{1} {\rm S}_{0}} $ & & & 1892 & 1956  \\
$ 3 \, {^{3} {\rm S}_{1}} $ & & & 1989 & 2035  \\
$ 3 \Delta {\rm SS} $ &  &  & 97 & 80  \\
\hline
$ 1 \, {^{1} {\rm P}_{1}} $ & $ K_{1} (1270) $
& 1273 $ \pm $ 7 & 1412 & 1397  \\
$
\begin{array}{c}
1 \, {^{3} {\rm P}_{2}} \\
1 \, {^{3} {\rm P}_{1}} \\
1 \, {^{3} {\rm P}_{0}}
\end{array}
$ &
$
\begin{array}{c}
K_{2}^{\ast} (1430) \\
K_{1} (1400) \\
K_{0}^{\ast} (1430)
\end{array}
$ &
$
\left.
\begin{array}{c}
1425.4 \pm 1.3 \\
1402 \pm 7 \\
1429 \pm 5
\end{array}
\right\}
1418
$ & 1412 & 1410  \\
\hline
$ 1 \, {^{1} {\rm D}_{2}} $ &
$ K_{2} (1580) $ &  & 1766 & 1775  \\
$
\begin{array}{c}
1 \, {^{3} {\rm D}_{3}} \\
1 \, {^{3} {\rm D}_{2}} \\
1 \, {^{3} {\rm D}_{1}}
\end{array}
$ &
$
\begin{array}{c}
K_{3}^{\ast} (1780) \\
K_{2} (1770) \\
K^{\ast} (1680)
\end{array}
$ &
$
\left.
\begin{array}{c}
1770 \pm 10  \\
1773 \pm 8  \\
1714 \pm 20
\end{array}
\right\}
1765
$ & 1766 & 1779  \\
\hline
$ 1 \, {^{1} {\rm F}_{3}} $ & & & 2046 & 2067  \\
$
\begin{array}{c}
1 \, {^{3} {\rm F}_{4}} \\
1 \, {^{3} {\rm F}_{3}} \\
1 \, {^{3} {\rm F}_{2}}
\end{array}
$ &
$
\begin{array}{c}
K_{4}^{\ast} (2045) \\
K_{3} (2320) \\
K_{2}^{\ast} (1980)
\end{array}
$ &
$
\left.
\begin{array}{c}
2045 \pm 9  \\
2324 \pm 24  \\
1975 \pm 22
\end{array}
\right\}
2130
$ & 2046 & 2069  \\
\hline
$ 1 \, {^{1} {\rm G}_{4}} $ & & & 2289 & 2317  \\
$
\begin{array}{c}
1 \, {^{3} {\rm G}_{5}} \\
1 \, {^{3} {\rm G}_{4}} \\
1 \, {^{3} {\rm G}_{3}}
\end{array}
$ &
$
\begin{array}{c}
K_{5}^{\ast} (2380) \\
K_{4} (2500) \\   \\
\end{array}
$ &
$
\left.
\begin{array}{c}
2382 \pm 14 \pm 19  \\
2490 \pm 20  \\
  \\
\end{array}
\right\}
$ & 2289 & 2320  \\
\hline
$ 1 \, {^{1} {\rm H}_{5}} $ & & & 2506 & 2545  \\
$
\begin{array}{c}
1 \, {^{3} {\rm H}_{6}} \\
1 \, {^{3} {\rm H}_{5}} \\
1 \, {^{3} {\rm H}_{4}}
\end{array}
$ &
$
\begin{array}{c}
   \\  \\   \\
\end{array}
$ &
$
\left.
\begin{array}{c}
    \\    \\   \\
\end{array}
\right\}
$ & 2506 & 2545  \\
\hline
\end{tabular}
\end{table}
\begin{table}
\vspace{1cm}
\centering
{Table IV. $ b \bar{b} $,
$ n_{f} = 4 $, $ \Lambda = 0.2 \; {\rm GeV} $.
(a)
truncation prescription
$ \bar\alpha_{\rm s} = 0.35 $,
$ \sigma = 0.2 \; {\rm GeV}^{2} $,
$ m_{b} = 4.763 \; {\rm GeV}  $.
(b)
Shirkov-Solovtsov
$ \alpha_{\rm s}( Q^{2} ) $,
$ \sigma = 0.18 \; {\rm GeV}^{2} $,
$ m_{b} = 4.898 \; {\rm GeV}  $.
\vspace{1mm}}
\begin{tabular}{ccccc}
\hline
States & & experimental values & (a) & (b)  \\
       &     &    (MeV)     &     (MeV)    &    (MeV)    \\
\hline
$ 1 \, {^{1} {\rm S}_{0}} $ & & & 9374 & 9374 \\
$ 1 \, {^{3} {\rm S}_{1}} $ & $ \Upsilon (1S) $
& 9460.30 $ \pm $ 0.26 & 9460 & 9460 \\
$ 2 \, {^{1} {\rm S}_{0}} $ & & & 9975 & 9988 \\
$ 2 \, {^{3} {\rm S}_{1}} $ & $ \Upsilon (2S) $
& 10023.26 $ \pm $ 0.31 & 10010 & 10023 \\
$ 3 \, {^{1} {\rm S}_{0}} $ & & & 10322 & 10342 \\
$ 3 \, {^{3} {\rm S}_{1}} $ & $ \Upsilon (3S) $
& 10355.2 $ \pm $ 0.5 & 10348 & 10368 \\
$ 4 \, {^{1} {\rm S}_{0}} $ & & & 10598 & 10618 \\
$ 4 \, {^{3} {\rm S}_{1}} $ & $ \Upsilon (4S) $
& 10580.0 $ \pm $ 3.5 & 10620 & 10639 \\
$ 5 \, {^{1} {\rm S}_{0}} $ & & & 10837 & 10854 \\
$ 5 \, {^{3} {\rm S}_{1}} $ & $ \Upsilon (10860) $
& 10865 $ \pm $ 8 & 10857 & 10872 \\
$ 6 \, {^{1} {\rm S}_{0}} $ & & & 11060 & 11070 \\
$ 6 \, {^{3} {\rm S}_{1}} $ & $ \Upsilon (11020) $
& 11019 $ \pm $ 8 & 11079 & 11089 \\
\hline
$ 1 \, {^{1} {\rm P}_{1}} $ & & & 9908 & 9918 \\
$
\begin{array}{c}
1 \, {^{3} {\rm P}_{2}} \\
1 \, {^{3} {\rm P}_{1}} \\
1 \, {^{3} {\rm P}_{0}}
\end{array}
$ &
$
\begin{array}{c}
\chi_{b2} (1P) \\
\chi_{b1} (1P) \\
\chi_{b0} (1P)
\end{array}
$ &
$
\left.
\begin{array}{c}
9912.6 \pm 0.5 \\
9892.7 \pm 0.6 \\
9859.9 \pm 1.0
\end{array}
\right\}
9900
$ & 9908 & 9920 \\
$ 2 \, {^{1} {\rm P}_{1}} $ & & & 10260 & 10279  \\
$
\begin{array}{c}
2 \, {^{3} {\rm P}_{2}} \\
2 \, {^{3} {\rm P}_{1}} \\
2 \, {^{3} {\rm P}_{0}}
\end{array}
$ &  
$
\begin{array}{c}
\chi_{b2} (2P) \\
\chi_{b1} (2P) \\
\chi_{b0} (2P)
\end{array}
$ &
$
\left.
\begin{array}{c}
10268.5 \pm 0.4 \\
10255.2 \pm 0.5 \\
10232.1 \pm 0.6
\end{array}
\right\}
10260
$ & 10260 & 10280 \\
\hline
\end{tabular}
\end{table}
\begin{table}
\centering
{Table V. $ c \bar{c} $,
$ n_{f} = 4 $, $ \Lambda = 0.2 \; {\rm GeV} $.
(a)
truncation prescription
$ \bar\alpha_{\rm s} = 0.35 $,
$ \sigma = 0.2 \; {\rm GeV}^{2} $,
$ m_{c} = 1.394 \; {\rm GeV}  $.
(b)
Shirkov-Solovtsov
$ \alpha_{\rm s}( Q^{2} ) $,
$ \sigma = 0.18 \; {\rm GeV}^{2} $,
$ m_{c} = 1.545 \; {\rm GeV}  $.
\vspace{1mm}}
\begin{tabular}{ccccc}
\hline
States & & experimental values & (a) & (b)  \\
       &     &    (MeV)     &     (MeV)    &    (MeV)    \\
\hline
$ 1 \, {^{1} {\rm S}_{0}} $ & $ \eta_{c} (1S) $
& 2979.7 $ \pm $ 1.5 & 2982 & 2977  \\
$ 1 \, {^{3} {\rm S}_{1}} $ & $ J/\psi (1S) $
& 3096.87 $ \pm $ 0.04 & 3097 & 3097  \\
$ 1 \Delta {\rm SS} $ &  & 117 & 115 & 119  \\
$ 2 \, {^{1} {\rm S}_{0}} $ & $ \eta_{c} (2S) $
& 3594 $ \pm $ 5 & 3575 & 3606  \\
$ 2 \, {^{3} {\rm S}_{1}} $ & $ \psi (2S) $
& 3685.96 $ \pm $ 0.09 & 3642 & 3670  \\
$ 2 \Delta {\rm SS} $ &  & 92 & 67 & 64  \\
$ 3 \, {^{1} {\rm S}_{0}} $ & & & 3974 & 4005  \\
$ 3 \, {^{3} {\rm S}_{1}} $ & $ \psi (4040) $
& 4040 $ \pm $ 10 & 4025 & 4054  \\
$ 4 \, {^{1} {\rm S}_{0}} $ & & & 4298 & 4323  \\
$ 4 \, {^{3} {\rm S}_{1}} $ & $ \psi (4415) $
& 4415 $ \pm $ 6 & 4341 & 4364  \\
\hline
$ 1 \, {^{1} {\rm P}_{1}} $ & & & 3529 & 3556 \\
$
\begin{array}{c}
1 \, {^{3} {\rm P}_{2}} \\
1 \, {^{3} {\rm P}_{1}} \\
1 \, {^{3} {\rm P}_{0}}
\end{array}
$ &
$
\begin{array}{c}
\chi_{c2} (1P) \\
\chi_{c1} (1P) \\
\chi_{c0} (1P)
\end{array}
$ &
$
\left.
\begin{array}{c}
3556.18 \pm 0.13 \\
3510.51 \pm 0.12 \\
3415.1 \pm 0.8
\end{array}
\right\}
3525
$ & 3530 & 3561 \\
$ 2 \, {^{1} {\rm P}_{1}} $ & & & 3925 & 3954 \\
$ 2 \, {^{3} {\rm P}} $ & & & 3927 & 3958 \\
\hline
$ 1 \, {^{1} {\rm D}_{2}} $ & & & 3813 & 3853 \\
$
\begin{array}{c}
1 \, {^{3} {\rm D}_{3}} \\
1 \, {^{3} {\rm D}_{2}} \\
1 \, {^{3} {\rm D}_{1}}
\end{array}
$ &
$
\begin{array}{c}
   \\ \psi (3836)  \\  \psi (3770)
\end{array}
$ &
$
\left.
\begin{array}{c}
   \\  3836 \pm 13  \\
3769.9 \pm 2.5
\end{array}
\right\}
$ & 3813 & 3854 \\
$ 2 \, {^{1} {\rm D}_{2}} $ & & & 4149 & 4183 \\
$
\begin{array}{c}
2 \, {^{3} {\rm D}_{3}} \\
2 \, {^{3} {\rm D}_{2}} \\
2 \, {^{3} {\rm D}_{1}}
\end{array}
$ &
$
\begin{array}{c}
   \\    \\   \psi (4160)
\end{array}
$ &
$
\left.
\begin{array}{c}
   \\    \\
4159 \pm 20
\end{array}
\right\}
$ & 4149 & 4184 \\
\hline
\end{tabular}
\end{table}
\begin{table}
\centering
{Table VI. Light-heavy quarkonium systems,
$ n_{f} = 4 $, $ \Lambda = 0.2 \; {\rm GeV} $. \\
(a)
truncation prescription
$ \bar\alpha_{\rm s} = 0.35 $,
$ \sigma = 0.2 \; {\rm GeV}^{2} $,
$ m_{b} = 4.763 \; {\rm GeV}  $,
$ m_{c} = 1.394 \; {\rm GeV}  $,
$ m_{u} =  m_{d} = 0.01 \; {\rm GeV}  $.
(b)
Shirkov-Solovtsov
$ \alpha_{\rm s}( Q^{2} ) $,
$ \sigma = 0.18 \; {\rm GeV}^{2} $,
$ m_{b} = 4.898 \; {\rm GeV}  $,
$ m_{c} = 1.545 \; {\rm GeV}  $,
$ m_{u}^{2} = m_{d}^{2} = 0.121 \, k - 0.025 \, k^{2} +
0.25 \, k^{4} $.
\vspace{1mm}}
\begin{tabular}{ccccc}
\hline
 States &  & experimental values & (a) & (b)  \\
    &   &  (MeV)  &  (MeV)  &  (MeV)   \\
\hline
$ q \bar{c} \; (q = u,d) $ &  &  &  &  \\
$ 1 \, {^{1} {\rm S}_{0}} $ &
$
\left\{
\begin{array}{c}
D^{\pm} \\
D^{0}
\end{array}
\right.
$
&
$
\left.
\begin{array}{c}
1869.3 \pm 0.5  \\
1864.5 \pm 0.5
\end{array}
\right\}
$
& 1875 & 1862  \\
$ 1 \, {^{3} {\rm S}_{1}} $ &
$
\left\{
\begin{array}{c}
D^{\ast}(2010)^{\pm} \\
D^{\ast}(2007)^{0}
\end{array}
\right.
$
&
$
\left.
\begin{array}{c}
2010.0 \pm 0.5  \\
2006.7 \pm 0.5
\end{array}
\right\}
$
& 2020 & 2012  \\
$ 1 \Delta {\rm SS} $ &  & 141 $ \pm $ 1 & 145 & 150  \\
$ 2 \, {^{1} {\rm S}_{0}} $ & $ D^{\prime} $ &
2580 & 2525 & 2615  \\
$ 2 \, {^{3} {\rm S}_{1}} $ & $ D^{\ast \prime} $ &
2637 $ \pm $ 8 & 2606 & 2683  \\
$ 2 \Delta {\rm SS} $ &  & 57 & 81 & 67  \\
\hline
$ 1 \, {^{1} {\rm P}_{1}} $ &  &  & 2474 & 2504  \\
$
\begin{array}{c}
\\
1 \, {^{3} {\rm P}_{2}} \\  \\
1 \, {^{3} {\rm P}_{1}} \\  \\
1 \, {^{3} {\rm P}_{0}}
\end{array}
$ &
$
\begin{array}{c}
\left\{
  \begin{array}{c}
    D_{2}^{\ast} (2460)^{\pm}  \\
    D_{2}^{\ast} (2460)^{0}
  \end{array}
\right.
\\
\left\{
  \begin{array}{c}
    D_{1} (2420)^{\pm}  \\
    D_{1} (2420)^{0}
  \end{array}
\right.
\\  \\
\end{array}
$ &
$
\left.
\begin{array}{c}
  \begin{array}{c}
    2459 \pm 4   \\
    2458.9 \pm 2.0
  \end{array}
\\
  \begin{array}{c}
    2427 \pm 5   \\
    2422.2 \pm 1.8
  \end{array}
\\  \\
\end{array}
\right\}
$ & 2475 & 2511  \\
\hline
$ q \bar{b} \; (q = u,d) $ &  &  &  &  \\
$ 1 \, {^{1} {\rm S}_{0}} $ &
$
\left\{
\begin{array}{c}
B^{\pm} \\
B^{0}
\end{array}
\right.
$
&
$
\left.
\begin{array}{c}
5278.9 \pm 1.8  \\
5279.2 \pm 1.8
\end{array}
\right\}
$
& 5273 & 5315   \\
$ 1 \, {^{3} {\rm S}_{1}} $ &
$ B^{\ast} $ & 5324.8 $ \pm $ 1.8 & 5339 & 5376   \\
$ 1 \Delta {\rm SS} $ &  & 46 $ \pm $ 3 & 66 & 61  \\
$ 2 \, {^{1} {\rm S}_{0}} $ &  &  & 5893 & 6034   \\
$ 2 \, {^{3} {\rm S}_{1}} $ &
$ B^{\ast \prime} $ & 5906 $ \pm $ 14 & 5933  & 6064  \\
\hline
$ 1 \, {^{1} {\rm P}_{1}} $ &  &  & 5791 & 5838  \\
$ 1 \, {^{3} {\rm P}} $ &  & 5825 $ \pm $ 14 & 5792 & 5841 \\
\hline
\end{tabular}
\end{table}
\begin{table}
\centering
{Table VII. Light-heavy quarkonium systems,
$ n_{f} = 4 $, $ \Lambda = 0.2 \; {\rm GeV} $. \\
(a)
truncation prescription
$ \bar\alpha_{\rm s} = 0.35 $,
$ \sigma = 0.2 \; {\rm GeV}^{2} $,
$ m_{b} = 4.763 \; {\rm GeV}  $,
$ m_{c} = 1.394 \; {\rm GeV}  $,
$ m_{s} = 0.2 \; {\rm GeV}  $.
(b)
Shirkov-Solovtsov
$ \alpha_{\rm s}( Q^{2} ) $,
$ \sigma = 0.18 \; {\rm GeV}^{2} $,
$ m_{b} = 4.898 \; {\rm GeV}  $,
$ m_{c} = 1.545 \; {\rm GeV}  $,
$ m_{s} = 0.381 \; {\rm GeV} $.
\vspace{1mm}}
\begin{tabular}{ccccc}
\hline
 States &  & experimental values & (a) & (b)  \\
    &   &  (MeV)  &  (MeV)  &  (MeV)  \\
\hline
$ s \bar{c} $ &  &  &  &  \\
$ 1 \, {^{1} {\rm S}_{0}} $ &
$ D_{s}^{\pm} $ & 1968.5 $ \pm $ 0.6 & 1982 & 1953  \\
$ 1 \, {^{3} {\rm S}_{1}} $ &
$ D_{s}^{\ast \pm} $ & 2112.4 $ \pm $ 0.7 & 2120 & 2102  \\
$ 1 \Delta {\rm SS} $ &  & 144 $ \pm $ 1 & 138 & 149  \\
$ 2 \, {^{1} {\rm S}_{0}} $ &  &  & 2617 & 2610  \\
$ 2 \, {^{3} {\rm S}_{1}} $ &  &  & 2698 & 2691  \\
\hline
$ 1 \, {^{1} {\rm P}_{1}} $ &  &  & 2547 & 2555  \\
$
\begin{array}{c}
1 \, {^{3} {\rm P}_{2}} \\
1 \, {^{3} {\rm P}_{1}} \\
1 \, {^{3} {\rm P}_{0}}
\end{array}
$ &
$
\begin{array}{c}
D_{sJ} (2573)^{\pm} \\
D_{s1} (2536)^{\pm} \\
\\
\end{array}
$ &
$
\left.
\begin{array}{c}
2573.5 \pm 1.7 \\
2535.35 \pm 0.34 \\
\\
\end{array}
\right\}
$ & 2548 & 2562  \\
\hline
$ s \bar{b} $ &  &  &  &  \\
$ 1 \, {^{1} {\rm S}_{0}} $ &
$ B_{s}^{0} $ & 5369.3 $ \pm $ 2.0 & 5364 & 5348  \\
$ 1 \, {^{3} {\rm S}_{1}} $ &
$ B_{s}^{\ast} $ & 5416.3 $ \pm $ 3.3 & 5429 & 5416  \\
$ 1 \Delta {\rm SS} $ &  & 47 $ \pm $ 4 & 65 & 68  \\
$ 2 \, {^{1} {\rm S}_{0}} $ &  &  & 5985 & 5969  \\
$ 2 \, {^{3} {\rm S}_{1}} $ &  &  & 6024 & 6009  \\
\hline
$ 1 \, {^{1} {\rm P}_{1}} $ &  &  & 5858 & 5857  \\
$ 1 \, {^{3} {\rm P}} $ &
$ B_{sJ}^{\ast} (5850) $ & 5853 $ \pm $ 15 & 5859 & 5860  \\
\hline
\end{tabular}
\end{table}

  As it is apparent the use of a running coupling
constant with infrared truncation does not improve essentially
the situation in comparison with the fixed coupling
constant case. The same happen with other prescription like the Dokshitzer
et al. prescription \cite{sanda}. On the contrary with the
Shirkov-Solovtsov coupling constant,
together with the running $u,d$ mass of eq. (\ref{eq:lqm}), the main
difficulties seem to be solved. Even the masses of $\pi$, $\eta_s$ and $K$
turn correct without destroying the agreement in the other part of the
spectrum. An exception could be represented by the masses of the $u\bar b$
channel that for some reason turn out to be
high. The use of an effective running mass even for $ m_{s} $
does not seem to
produce any substantial modification. With reference to the mass of
$\eta_s$ reported as a data in fig. \ref{figuussus}
concerning the $s \bar s$ sector, it should be
mentioned that the quantity is derived from the experimental values of
$\eta$ and $\eta^\prime$ masses with conventional assumptions on the
origin of this particles as a mixing of ${1\over \sqrt{2}}(u \bar u +
d \bar d)$ and $s \bar s$ states.


\section{Conclusions}

In conclusion, starting from a reasonable ansatz on the Wilson loop
it is possible to develop a {\it second order} Bethe-Salpeter approach
to the $q \bar q$ bound state problem. From this, by a standard reduction
technique, an eigenvalue equation
for a mass squared operator $M^2=(w_1+w_2)^2 + U $
is obtained, that can be effectively applied to the calculation of the
spectrum.

  If only the one gluon exchange contribution is included in the
perturbative part of the Bethe-Salpeter kernel, but a truncated running
constant $\alpha_{\rm s}(Q^2)$ is used, a reasonable good reproduction of
the data is already obtained (as far as calculable) with the relevant
exception of the light pseudo scalar mesons. However if the
Shirkov-Solovtsov prescription for $\alpha_{\rm s}(Q^2)$ is adopted and
a phenomenological running mass is used for the $u$ and $d$ quarks, even
the light pseudo scalar mesons occur in the correct range.

    An open problem is a resolution of the complicated Dyson-Schwinger
equation in some reasonable approximation (\ref{eq:dshom}) and an actual
derivation of an equation of the type (\ref{eq:lqm}) from first principle.



\begin{thebibliography}{99}
%
\bibitem{bmp}
  N. Brambilla, E. Montaldi, G.M. Prosperi, {\sl Phys. Rev.}
  {\bf D 54} (1996) 3506;
  G.M. Prosperi, 
  in {\sl Problems of Quantum Theory of Fields}, Pag. 381,
  B.M. Barbashov, G.V. Efimov, A.V. Efremov Eds.
  JINR Dubna 1999,
  hep-ph/9906237.
\bibitem{simon}
  For a different approach: Yu.S. Kalashnikova, A.V. Nefediev, Yu.A. Simonov,
  {\sl Phys. Rev.} {\bf D 64} (2001) 014037;
  Yu.S. Kalashnikova, A.V. Nefediev,
  {\sl Phys. Lett.} {\bf B 530} (2002) 117,
  and references therein.
\bibitem{quadratic}
  M. Baldicchi, G.M. Prosperi, {\sl Phys. Rev.} {\bf D 62}
  (2000) 114024;
  {\sl Fizika} {\bf B 8} (1999) 2, 251;
  M. Baldicchi, in
  {\sl ``QCD: Perturbative or Nonperturbative ?''}
  Pag. 325,
  L. S. Ferreira, P. Nogueira, J. I.Silva-Marcos Eds.
  World Scientific (2001), hep-ph/9911268.
\bibitem{sanda}
  A.I. Sanda {\sl Phys. Rev. Lett.}
  {\bf 42} (1979) 1658;
  Yu.L. Dokshitzer, A. Lucenti, G. Marchesini, G.P. Salam,
  {\sl JHEP} 9805 (1998) 003;
  Yu.L. Dokshitzer, in $ 29^{th} $ {\sl International
  conference on High-Energy Physics} (ICHE 98), Vancouver, Canada,
  A. Astbury, D. Axen, J Robinson Eds.
  World Scientific (1999), hep-ph/9812252;
  Yu.L. Dokshitzer, V.A. Khoze, S.I. Troyan,
  {\sl Phys. Rev.} {\bf D 53} (1996) 89.
\bibitem{shirkov}
  D.V. Shirkov, I.L. Solovtsov, {\sl Phys. Rev. Lett.}
  {\bf 79} (1997) 1209;
  {\sl Theor. Math. Phys.} {\bf 120} (1999) 1220;
  K.A. Milton, I.L. Solovtsov, O.P. Solovtsova,
  {\sl Phys. Rev.} {\bf D 64} (2001) 016005;
  N.G.Stefanis, W. Schroers, H.-Ch. Kim, {\sl Phys. Lett.}
  {\bf B 449} (1999) 299,
  see also {\sl Eur. Phys. J.} {\bf C 18} (2000) 137;
  A. V. Nesterenko, {\sl Phys. Rev.} {\bf D 62} (2000) 094028;
  {\bf D 64} (2001) 116009;
  A. V. Nesterenko, {\sl Mod. Phys. Lett.}
  {\bf A 15} (2000) 2401;
  A. V. Nesterenko, I.L. Solovtsov, {\sl ibid.}
  {\bf A 16} (2001) 2517;
  S. J. Brodsky, S. Menke, C. Merino,
  {\sl Phys. Rev.} {\bf D 67} (2003), 055008.
\bibitem{infrared}
  M. Baldicchi and G. M. Prosperi {\sl Phys. Rev.}
  {\bf D 66} (2002) 074008.
\bibitem{chiral}
  M.B. Hecht, C.D. Roberts, S.M. Schmidt,
  {\sl Phys. Rev.} {\bf C 63} (2001) 025213, and references therein.
\bibitem{barch}
  A. Barchielli, E. Montaldi, G.M. Prosperi,
  {\sl Nucl. Phys.} {\bf B 296} (1988) 625;
  Erratum-ibid. {\bf B 303} (1988) 752;
  A. Barchielli, N. Brambilla, G.M. Prosperi,
  {\sl Il Nuovo Cimento} {\bf 103 A} (1990) 59;
  N. Brambilla, P. Consoli, G.M. Prosperi,
  {\sl Phys. Rev.} {\bf D 50} (1994) 5878.
\bibitem{data}
The Review of Particle Physics, K. Hagiwara {\it et al.},
{\sl Phys. Rev.} {\bf D 66} 010001 (2002).

\end{thebibliography}
\end{document}